\documentstyle[a4,epsf,12pt]{article}
\def\plb#1{Phys.~Lett.~{\bf B#1}}
\def\npb#1{Nucl.~Phys.~{\bf B#1}}
\def\prl#1{Phys.~Rev.~Lett.~{\bf #1}}
\def\prd#1{Phys.~Rev.~{\bf D#1}}

\def\btodands{$B_{(s)}\to D_{(s)}$}
\def\btodsands{$B_{(s)}\to D^*_{(s)}$}
\def\btod{$B\to D$}
\def\btods{$B\to D^*$}
\def\bksg{B\to K^*\gamma}
\def\btodsl{B\to D^*\ell\nu}
\def\btodl{B\to D\ell\nu}
\def\bstodssl{B_s\to D_s^*\ell\nu}
\def\bstodsl{B_s\to D_s\ell\nu}
\def\lqcd{\Lambda_{\rm QCD}}
\def\as{\alpha_s}
\def\kc{\kappa_{crit}}

\def\wid#1{\Gamma\l(#1\r)}
\def\br#1{{\cal BR}\l(#1\r)}

\def\stat{({\rm stat.})}
\def\syst{({\rm syst.})}

\def\su4sf{$SU(4)_{spin\times flavor}$}

\def\tab#1{Table~\ref{#1}}

\def\fig#1{Fig.~\ref{#1}}
\def\figs#1#2{Figs.~\ref{#1} and \ref{#2}}
\def\sec#1{Section \ref{#1}}
\def\eq#1{Eq.~(\ref{#1})}
\def\eqs#1#2{Eqs.~(\ref{#1}) and (\ref{#2})}

\def\s#1{{\bf #1}}
\def\d{\delta}
\def\g{\gamma}
\def\k{\kappa}
\def\gm{\g_\mu}
\def\e{\epsilon}
\def\a{\alpha}
\def\b{\beta}
\def\w{\omega}
\def\L{\Lambda}
\def\la{\langle}
\def\ra{\rangle}
\def\Ds{D\hskip -0.27cm\slash}
\def\l{\left}
\def\r{\right}
\def\ord#1{{\cal O}\l(#1\r)}
\def\toas#1{\buildrel\;\;#1\;\;\over\longrightarrow}

\newcommand{\beq}{\begin{equation}}
\newcommand{\eeqa}{\end{eqnarray}}
\newcommand{\beqa}{\begin{eqnarray}}
\newcommand{\eeq}{\end{equation}}

\newcommand{\mev}{{\,\rm MeV}}
\newcommand{\gev}{{\,\rm GeV}}

\newcommand{\plus}{\makebox[15pt][c]{$+$}}
\newcommand{\minus}{\makebox[15pt][c]{$-$}}
\newcommand{\figurebox}[2]{\fbox{\vbox to#2in{\hbox to #1in{\hfil} \vfil}}}
\newcommand{\errr}[2]{\raisebox{0.08em}{\scriptsize {$\;\begin{array}{@{}l@{}}
                          \plus\makebox[1.35em][r]{#1} \\[-0.12em] 
                          \minus\makebox[1.35em][r]{#2} 
                        \end{array}$}}}
\newcommand{\err}[2]{\raisebox{0.08em}{\scriptsize {$\;\begin{array}{@{}l@{}}
                          \plus\makebox[0.95em][r]{#1} \\[-0.12em] 
                          \minus\makebox[0.95em][r]{#2} 
                        \end{array}$}}}
\newcommand{\er}[2]{\raisebox{0.08em}{\scriptsize {$\;\begin{array}{@{}l@{}}
                          \plus\makebox[0.55em][r]{#1} \\[-0.12em] 
                          \minus\makebox[0.55em][r]{#2} 
                        \end{array}$}}}

\def\BorderBox#1#2{\vbox{
                               #2%
			}}%
\def\InsertFigure[#1 #2 #3 #4]#5#6#7{%
	\epsfxsize=#6%
	\epsfysize=#7%
	\epsfbox[#1 #2 #3 #4]{#5}%
	}
\newcounter{myfigure}

\newlength{\captionwidth}
\newcommand{\ewxy}[2]{\setlength{\epsfxsize}{#2}\epsfbox[100 30 490 490]{#1}}
\newcommand{\ewxydiag}[2]{\setlength{\epsfxsize}{#2}\epsfbox[0 300 640 590]{#1}}
\begin{document}


\begin{titlepage}

\begin{flushright}
CPT-94/P.3098\\
Southampton Preprint SHEP 94/95-07\\
\end{flushright}

\vspace*{5mm}

\renewcommand{\thefootnote}{\fnsymbol{footnote}}

\begin{center}
{\Huge Weak Decays of $B$ Mesons and Lattice QCD
\footnote{Based on an invited lecture given at the XXXIV$^{th}$ Cracow
School of Theoretical Physics, Zakopane, Poland, 
May 31 to June 10, 1994. To be published in the proceedings.}}
\\[15mm]
{\large Laurent LELLOUCH
\footnote{{\it e-mail address: lellouch@cpt.univ-mrs.fr}
}}\\[3mm]
Centre de Physique Th\'eorique
\footnote{Unit\'e Propre de Recherche 7061}\\
CNRS--Luminy, Case 907\\
F-13288 Marseille Cedex 9, France\\
\renewcommand{\thefootnote}{\fnsymbol{footnote}}
\setcounter{footnote}{0}

\end{center}
\vspace{5mm}
\begin{abstract}
After a pedagogical introduction to the calculation of weak matrix
elements on the lattice, I review some of the lattice's most recent
predictions concerning the weak decays of $B$-mesons. Amongst the
topics covered are the determinations of the leptonic decay constant
$f_B$, of the form factors relevant for semi-leptonic $B\to
D(D^*)\ell\bar\nu$ decays and of the hadronic matrix element which
describes the rare decay $B\to K^*\g$.  Emphasis is placed on the
results of the UKQCD Collaboration.  Corrections to the heavy-quark
limit are discussed extensively.
\end{abstract}

\end{titlepage}



\section{Introduction}
\label{intro}
Weak decays of hadrons are a very rich source of information about the
Standard Model. Not only do they enable us to determine from
experiment the parameters which are associated with the flavor sector
of the Standard Model but they also provide a rich testing ground for
understanding the non-perturbative dynamics of the strong interaction.
These two aspects of hadronic, weak decays are in fact inseparable. To
be more specific, the Standard Model prediction for the rate of such a
decay follows the pattern
\beq
\mbox{rate} = \left\{
\begin{array}{l}
\mbox{kinematical}\\
\mbox{factors}\\
\end{array}
\right\}
\left\{
	\begin{array}{l}
	\mbox{Non-Perturbative}\\
	\mbox{QCD factor}\\
	\end{array}
	\right\}
\left\{
	\begin{array}{l}
	\mbox{CKM}\\
	\mbox{factor}\\
	\end{array}
	\right\}
\eeq
from which it is clear that in order to extract the
Cabibbo-Kobayashi-Maskawa (CKM) factor from an
experimental measurement of the rate, one has to understand or at
least calculate the non-perturbative QCD factor. The lattice
formulation of QCD, together with large scale numerical simulations,
provides a natural framework for evaluating this factor.  It is in
fact the only systematic, first-principle approach we know for
quantifying non-perturbative, strong-interaction dynamics with the QCD
lagrangian as a starting point. This should not be taken to mean that
it is only approach or even the most efficient approach, as QCD
sum-rules, low energy effective theories and quark models have
provided many important and reliable results throughout the years.

The latest chapter in the story of hadronic weak decays concerns
hadrons containing a $b$-quark. The decays of these hadrons are
interesting phenomenologically because their study will enable us to
determine the least well know column of the CKM matrix. They are also
interesting theoretically because they provide a ground for testing
heavy-quark symmetry. Heavy-quark symmetry is a symmetry of QCD that
arises in the limit that the mass of the heavy quark is much larger
than the QCD scale $\lqcd$. In that limit one finds that the dynamics
of the light quarks and gluons coupled to the heavy quark become
independent of the heavy quark's flavor and spin. To the extent, then,
that $\lqcd$ is negligible compared to the masses of the charm and
beauty quarks, QCD exhibits a new $SU(4)_{spin\times flavor}$ symmetry
which acts on the multiplet
$(c\uparrow,c\downarrow,b\uparrow,b\downarrow)$\cite{russians,iw}.  This
symmetry tremendously simplifies the description of the decays of
hadrons containing a heavy quark.
It has in fact been incorporated into the framework of an effective
theory\cite{hqet} known as Heavy Quark Effective Theory (HQET). This
makes it possible to systematically calculate corrections to the symmetry
limit order by order in inverse powers of the heavy-quark masses
much in the same way that Chiral Lagrangians enables one to calculate
corrections to the predictions of the spontaneously broken chiral symmetry
of QCD in the light quark sector.\footnote{For a
comprehensive review of HQET and heavy-quark symmetry see the lectures
of K.~Zalewski in this volume or the review of \cite{bible}.} I would
like to emphasize, at this point, that the lattice is supremely well
suited for studying the range of applicability of heavy-quark symmetry,
for the masses of the heavy quarks are
free parameters in lattice calculations and the dependence of
results on heavy-quark mass can be studied in detail.

In the present article I will review some of the latest results of
lattice calculations of weak matrix elements of beautiful hadrons.  I
will concentrate mainly on the results of the UKQCD Collaboration
because this article is not meant to be thorough and systematic review
of the subject but rather an introduction to the methods,
possibilities and limitations of this rapidly growing field. In
\sec{latcomp} I will provide a introduction to lattice calculations of
weak matrix elements. For a more complete review, see the lectures of
R.~Gupta in the present volume or the reviews in \cite{reviews}. For
reviews about $b$-physics on the lattice, see Ref.\cite{brev}. In
\sec{fb}, I will present results for the $B$-meson decay constant,
$f_B$, and will describe what these results teach us about heavy-quark
symmetry. I will then turn to the subject of semi-leptonic \btodands\
and \btodsands\ decays in \sec{semi-lep}. I will present
determinations of the Isgur-Wise functions $\xi_{u,d}$ and $\xi_s$
which parametrize the strong interaction effects to leading order in
heavy-quark mass in these decays. I will show how the function
$\xi_{u,d}$ can be used to extract the CKM matrix-element $V_{cb}$
from an experimental measurement of the differential decay rate for
$\btodsl$ decays. I will also present many tests of the heavy-quark
symmetry which involve comparing \btod\ and \btods\ decays for many
different values of the $b$ and $c$ quark masses. \sec{bsgsec} will be
dedicated to a lattice evaluation of the rate for the rare process
$\bksg$. This process occurs through a
flavor-changing-neutral-current, penguin diagram and was first
observed in 1993 by the CLEO collaboration\cite{cleobksg}. I will end
the article in \sec{concl} with a summary and conclusions.



\section{Matrix Element Calculations on the Lattice}
\label{latcomp}


\subsection{What We Calculate}
\label{what}
To obtain matrix elements in lattice calculations we compute
expectation values of products of gluon and quark fields at different
spacetime points. We do so using the path integral formulation of QCD
in euclidean spacetime. Thus, given a product ${\cal
O}(x_1,\cdots,x_n)$ of quark and gluon fields, we compute $n$-point
functions:
\beq
\la {\cal O}(x_1,\cdots,x_n)\ra=\frac{1}{Z}\int dA_\mu d\bar\psi d\psi\ 
{\rm e}^{-S_{QCD}} {\cal O}(x_1,\cdots,x_n)
\ ,
\label{npoint}
\eeq
where $S_{QCD}$ is the QCD action and
\beq
Z=\int dA_\mu d\bar\psi d\psi\ 
{\rm e}^{-S_{QCD}} 
\ .
\label{partfn}
\eeq

The simplest example of an
$n$-point function is the 2-point function. To determine the decay
constant of a $B$-meson, for instance, we would evaluate the
2-point function
\beq
\sum_{\s x}\la\bar q\g^\mu\g^5b(\s x,t)\ \bar b\g^5q(0)\ra
\ ,
\label{fb2pt}
\eeq
where $q$ is a light-quark field and the sum over $\s x$ ensures that
the meson is at rest. If we insert a complete set of states between
the two operators in \eq{fb2pt} we find that this correlator
decays exponentially in time in the limit of large $t$. 
The rate of the decay is governed by the mass of ground state and
its amplitude is proportional to the weak matrix element,
$\la 0|\bar q\g^\mu\g^5b(0)|B\ra$, we are
after:
\beq
\sum_{\s x}\la\bar q\g^\mu\g^5b(\s x,t)\ \bar b\g^5q(0)\ra
\toas{t\to\infty}
\frac{\la 0|\bar q\g^\mu\g^5b(0)|B\ra\,
\la B|\bar b\g^5q(0)|0\ra}{2m_B}\ 
{\rm e}^{-m_Bt}
\ .
\label{2ptlim}
\eeq

The next level of complexity is the 3-point function. The matrix element
$\la D(\s p')|\bar c\g^\mu b(0)|B(\s p)\ra$, for instance, is obtained
from the 3-point function
\beq
\sum_{\s x,\s y}{\rm e}^{-i\s p'\cdot\s y-i\s q\cdot\s x}
\la\bar c\g^5 q(\s y,t_f)\,\bar c\g^\mu b(\s x,t)\,\bar b\g^\mu q(0)\ra
\label{btod3pt}
\eeq
where $\s q=\s p-\s p'$. In the limit that
$t,\,t_f-t\to\infty$, this 3-point function reduces to
\beq
\frac{\la 0|\bar q\g^5b(0)|D(\s p')\ra\,
\la D(\s p')|\bar c\g^\mu b(0)|B(\s p)\ra\,
\la B(\s p)|\bar b\g^5q(0)|0\ra}{4E_DE_B}
\ {\rm e}^{-E_Bt-E_D(t_f-t)}
\ ,
\label{btod3ptlim}
\eeq
where $E_B$ and $E_D$ are the energies of the $B$ and $D$ mesons,
respectively.



\subsection{How We Calculate}
\label{how}

The first step in evaluating the path integral of \eq{npoint} is to
approximate spacetime by a finite, hypercubic lattice.  Quark fields
are then placed on the sites of this lattice and the gluon fields on
the links between these sites.  This reduces the integral over gluon
and quark fields to an integral over a finite number of degrees of
freedom which makes it amenable to numerical methods. It is important
to note, at this point, that to preserve gauge invariance on a
discretized spacetime one works with elements of the group--the link
variables $U_\mu(x)=e^{iagA_\mu(x)}$--instead of with the gauge fields
$A_\mu(x)$ (see Ref.\cite{wilsonorig}, for example). 
\footnote{Here $g$ is the gauge coupling
constant and $a$, the lattice spacing. This lattice spacing acts
as a non-perturbative regulator: modes with momenta larger 
than $\pi/a$ are absent regardless of whether the theory is
used perturbatively or non-perturbatively.} 
This means, for instance, that the integrals over gauge
fields in \eqs{npoint}{partfn} are actually integrals over the compact
link variables $U_\mu(x)$.

As a first step in the evaluation of this lattice path integral, we
generate a number $N$ of gluon configurations by a Monte-Carlo
simulation\cite{mc} with the probability distribution
\beq
{\cal P}\equiv \frac{{\rm e}^{-S_{gluon}}\times {\rm det}(\Ds+M)}{Z}
\ ,
\label{dist}
\eeq
where $S_{gluon}$ is a discretized gluon action\cite{wilsonorig},
$D_\mu$ a discretized $SU(3)$ covariant derivative in the fundamental
representation and $M$, the quark-mass matrix.  Then, for example, to
obtain the propagator of a $B$-meson,
\beq
\la \bar u\g^5b(x)\,\bar b\g^5u(0)\ra
\ ,
\label{bmesprop}
\eeq
we use Wick's theorem to reduce the quark bilinears in
\eq{bmesprop} to a product of quark propagators and do so for each one of
the $N$ gluon backgrounds. The meson propagator then becomes the
following average of a product of quark propagators and Dirac matrices over
gluon backgrounds,
\beq
\la \bar u\g^5b(x)\,\bar b\g^5u(0)\ra=\frac{1}{N}
\sum_{U_\mu} {\rm Tr}\l(\g^5S_b(0,x;U_\mu)\g^5S_u(x,0;U_\mu)\r)
\ ,
\label{bmesproptr}
\eeq
which can be evaluated numerically once the propagators,
$S_{b(u)}(0,x;U_\mu)$, for the $b$ and $u$ quarks in the gluon
background $U_\mu$ have been calculated using matrix inversion
algorithms such as the conjugate gradient algorithm\cite{matinv}.



\subsection{Limitations and Sources of Errors}
\label{limitations}

Even though the numerical evaluation of the path integral of
\eq{npoint} described in \sec{how} 
is an ab initio calculation starting with the QCD lagrangian, the
final results have errors which are due to the approximations that
have to be made. These errors can be classified into two broad
categories.  There are {\it statistical errors} which arise because we
use statistical methods to evaluate the path integral and {\it
systematic errors} which arise because we approximate spacetime by a
finite lattice of points and because we usually have to neglect the
contributions of fermion loops. One may wonder, at this point, what
advantage the lattice has over other means of evaluating hadronic
properties.  The difference lies in that the errors made in a lattice
calculation can be reduced systematically by increasing the number of
configurations and the physical volume of the lattice, by
decreasing the lattice spacing or by designing lattice actions
which converge to the continuum limit more rapidly.

\subsubsection{Statistical Errors}
Statistical errors arise because we approximate the integral over
gauge fields, which is an integral over an infinite number of
configurations, by a sum over a finite number, $N$, of
configurations. According to the central limit theorem, the error made
on $n$-point functions is proportional to $1/\sqrt{N}$ in the
limit of large $N$, so that statistical errors decrease as the number
of configurations increases. Typical values of $N$ in QCD
calculations are on the order of 100 and the corresponding errors are
on the order of a few percent for many quantities. We evaluate
these errors using standard statistical methods such as the bootstrap
or jacknife method\cite{EFRON}. These methods enable us to
approximately determine the distribution of results we would find were we
to repeat our $N$-configuration calculation many times. The statistical
error we quote on our result is chosen to include to the central 68\% 
of this distribution. 

\subsubsection{Discretization Errors}
Discretization errors are due to the fact that we approximate
spacetime by a discrete lattice of points. To see how these errors
arise, consider the symmetric lattice derivative of a function, $f$,
of spacetime in the limit that the lattice spacing $a$ is taken to
zero:
\beq
\frac{f(x+a\hat\mu)-f(x-a\hat\mu)}{2a}\toas{a\to 0}
\partial_\mu f(x)+\ord{a^2}
\ ,
\label{latder}
\eeq
where $\hat\mu$ is a unit vector in
the $\mu$-direction. It is clear from \eq{latder} that approximating
the derivative of $f$ by a finite difference introduces an error
of order $a^2$. Before exploring these discretization errors
in more detail, though, I shall first describe how we determine the lattice
spacing $a$ from our simulations.

\bigskip
{\it Setting the Scale}

As you well know, the bare coupling constant $g(a)$ is related to the
cutoff $a$ through dimensional transmutation.  Before
performing our simulation, however, we only know what this
relation is through perturbation theory up to some
finite order in $g(a)$. So it is preferable not to fix both quantities from
the start. What we do is pick a value for the bare coupling,
$\beta=2N_c/g^2(a)$,
\footnote{Here, perturbation theory is a useful guide for it tells us
approximately what value of $\beta$ yields the desired lattice spacing.}  
and rewrite the QCD lagrangian in terms of
dimensionless quantities, measured in units of the lattice spacing
$a$. All explicit dependence on $a$ then disappears from
the problem and $a$ need not be fixed before performing the
simulation. Once the simulation has been performed, however, $a$ must
be determined so that physical dimensions may be restored. 
To do this, we pick a dimensionful, physical
quantity such as the string tension $\sigma$. The lattice
spacing is then the ratio of the experimental value, $\sigma^{\rm expt}$,
to the dimensionless quantity,
$\sigma^{\rm latt.}$, that our simulation yields, raised to the appropriate
power:
\beq
a^{-1}=\sqrt{\frac{\sigma^{\rm expt}}{\sigma^{\rm latt}}}
\ .
\label{spacing}
\eeq

To set the scale any dimensionful quantity will do though different 
quantities will lead to more or less accurate determinations
of the lattice spacing.
\footnote{For quantities that depend on quark masses, one has
to first determine the quark's bare mass.} In \tab{scale}, I show the
values for $a^{-1}$ that our collaboration obtains from various physical
quantities. All of these quantities should, in principle, yield the
same lattice spacing assuming, of course, that QCD is correct.  The
reason why the values in \tab{scale} do not agree within the quoted
statistical errors is because we also make systematic errors. In fact,
the range of values that we obtain for $a^{-1}$ is a good indication
of the size of the systematic errors we make.
\begin{table}
\centering
\begin{tabular} {|c|c|}\hline
Physical Quantity  & $a^{-1}$ ($\gev$)\\
\hline
$\sqrt{\sigma}$ (string tension) & 2.73(5) \\
$m_\rho$ (mass of the $\rho$) & $2.7\er{1}{1}$ \\
$m_N$ (mass of the nucleon) &$3.0\er{2}{3}$ \\
$m_\Delta$ (mass of the $\Delta$) & $2.5\er{2}{2}$ \\
\hline
\end{tabular}
\begin{center}
\begin{minipage}[t]{\captionwidth}
\caption{Values of the inverse lattice spacing, $a^{-1}$, for 
UKQCD's $\beta=6.2$,
$24^3\times48$ lattice as obtained from different physical quantities
\protect{\cite{strange}}.
Ratios of these different values indicate how well the corresponding
ratio of physical quantities is predicted by the simulation. 
$a^{-1}(m_\rho)/a^{-1}(m_N)$, for instance, indicates that the
ratio $m_N/m_\rho$ comes out 10\% low in our simulation but
with errors large enough to make it consistent with experiment.
\label{scale}}
\end{minipage}
\end{center}
\end{table}

\bigskip
Having established how one determines the lattice spacing, we 
return to the discussion of discretization errors and how to
reduce them. The reason why
these errors are particularly important in simulations of hadrons
containing a heavy quark is because these quarks have very short
compton wavelengths: an accurate description of their
quantum propagation appears to require a very fine lattice. There
are in fact many different ways to describe heavy quarks on a lattice.
These different approaches fall into two broad categories. In the first
category, the quark action used is simply a discretization of the
full continuum quark action. In the second, that of effective theories,
this discretized action
is further expanded in inverse powers of the heavy-quark mass as
well as in powers of the heavy quark's velocity if the heavy quark
is non-relativistic.

\medskip
{\it The Wilson Action}

One of the standard discretizations of the continuum quark action
is the Wilson action\cite{wilsonorig}:
\beqa
S^W=a^4\ \sum_x & \Biggl\{-\frac{1}{2a}\sum_\mu
\l[\bar q(x)(r-\g_\mu)U_\mu(x)q(x+a\hat\mu)
+\bar q(x+a\hat\mu)(r+\g_\mu)U_\mu^\dagger(x)q(x)\r]\nonumber\\
&+\bar q(x)\l(m_o+\frac{4r}{a}\r)q(x)\Biggr\}
\ ,
\label{sw}
\eeqa
where $r$ is a number (typically taken to be 1) known as the
Wilson parameter. 
In the naive continuum limit, this action reduces to
\beq
S^W\toas{a\to 0} \int\bar q\l(\Ds+m_o\r)q
-\frac{ar}{2}\int\bar q D^2q+\ord{a^2}
\ .
\label{swcont}
\eeq
Because of the $\ord{a}$ term, we expect the leading discretization
errors, in a simulation of a hadron containing a heavy quark, to be on
the order of $am_Q$, where $m_Q$ is the mass of the heavy quark.  For
these errors to remain under control, we must require $am_Q\ll 1$.
Since typical values of the inverse lattice spacing are on the order
of $3\gev$, the $b$ quark, which has a mass of about $4.5\gev$, cannot
be simulated directly. Thus, the strategy used to obtain properties of
$b$-hadrons with the Wilson action is to calculate these properties
for a collection hadrons whose heavy quarks have a mass in the range of
the charm quark mass and then to extrapolate these properties in
heavy-quark mass to $m_b$, using HQET as a guide. But even for the
charm quark, we expect large discretization errors as $am_c\sim 0.4$
for a typical value of the lattice spacing. So if the Wilson action is
to be used, one has to find a way to reduce these discretization
errors.

The brute force approach to reducing discretization errors is to
increase the number of lattice sites and reduce the lattice
spacing. This requires ever faster computers. The second type of approach
is to design lattice theories which converge
faster to the continuum limit. It is to this second approach
we now turn.

\medskip
{\it Improving the Wilson Action: Normalization of Quark Fields} 

The impetus for the improvement that I am going to describe comes from an
analysis of the free-fermion propagator on the lattice. One finds,
using the action given in \eq{sw}, that this propagator is given, at zero
momentum, by
\beq
\sum_{\s x}S^W(x,0)={\rm e}^{-am}\ \l[\frac{1+\g_o}{2}\,{\rm e}^{-mt}\r]
\ ,
\label{freeprop}
\eeq
where the term in brackets is the continuum, free propagator for
a fermion of mass
\beq
am={\rm ln}\l(1+am_o\r)
\ .
\label{logam}
\eeq
In the limit that $am_o$ vanishes, the lattice propagator in
\eq{freeprop} reduces to the continuum propagator for a fermion of mass
$m_o$, as it should. However, when $am_o\sim 1$, as it is for the
charm, this lattice propagator starts deviating from continuum
behaviour.  To remedy this problem, it has been
suggested\cite{lepagehq,kronmac} that one should let $m$, defined in
\eq{logam}, be the bare quark mass and that one should rescale the
lattice quark field $q$ according to
\beq
q\longrightarrow {\rm e}^{am/2} q
\ .
\label{rescq}
\eeq
For the charm quark, this corresponds to rescaling the quark field by
a factor of about 1.2, quite a large factor.
 
Even though this procedure clearly improves the free lattice
propagator it must be stressed that it is not a systematic procedure
and that it is unclear, at this point, what the remaining discretizations 
are once interactions are switched on. For instance, one would expect
that $\a_s am_o$ and $a\lqcd$ discretization errors are still present.

It has even been suggested recently\cite{kronmac} that it is
possible to use the Wilson action for any value of $am_o$--even
$am_o\gg 1$--as long as one performs the rescaling of \eq{rescq} and one
defines the mass, M, of the simulated hadron as
\beq
M\equiv \frac{1}{2}\l(\frac{\partial E}{\partial\s p^2}
\r)_{\s p^2=0}^{-1}
\ .
\eeq
Here again, however, one is entitled to suspect that large $\a_s am_o$-errors
might contaminate results for matrix elements.

\medskip
{\it Improving the Wilson Action: Symanzik's Procedure}

A more systematic approach to reducing discretization errors was
initiated by Symanzik in the early eighties\cite{symanzik}. Here the
idea is to formally remove discretization errors order by order
in $a$ by introducing higher-order operator corrections to the
action. The first step in this program, as it applies to
lattice QCD, was carried out by Sheikholeslami and Wohlert who
proposed the $\ord{a}$-improved action\cite{sw}
\beq
S^{SW}=S^W-irg_oa^4\sum_{x,\mu,\nu}\bar q\,\sigma_{\mu\nu}P_{\mu\nu}\,q(x)
\ ,
\label{clover}
\eeq
where $P_{\mu\nu}$ is a discretization of the field-strength tensor
$F_{\mu\nu}$. This action is known as the Sheikholeslami-Wohlert
action or clover action. The name ``clover'' comes from the fact that
a possible expression for $P_{\mu\nu}$ is the four-leafed-clover
product of the four plaquettes that originate at $x$ and lie in the
$\mu\nu$-plane.

It has been shown\cite{improv} that when this action is used
in conjunction with the field rotation
\beq
q\longrightarrow q^R\equiv\l(1-\frac{ar}{2}\Ds\r)q
\label{rot}
\eeq
the discretization errors made on matrix elements are reduced from
$\ord{a}$ to $\ord{\as a}$, where $\as$ is the strong coupling
constant. Thus, the discretization errors which we evaluated to be on
the order of $am_c\sim 40\%$ when simulating the charm with the
Wilson action on a typical present day lattice (i.e. $\beta=6.2)$ are
reduced to about $\as am_c\sim 4\%$ when the clover action is used
\footnote{The value of $\as$ used here is the ``boosted'' value
which incorporates the effects of tadpole diagrams
and is therefore more physical. For the UKQCD simulation at $\beta=6.2$
it is $\as=3/(2\pi\beta u_o^4)$, where $u_o$ is a measure of the 
expectation value of a link variable which we take to be $u_o=1/(8\kc)$.
$\kc=0.14315$ is the critical value of the hopping parameter.}.

\medskip
{\it Improving Lattice Actions: The Perfect Action}

A little over a year ago P.~Hasenfratz and F.~Niedermayer\cite{peter}
raised the tantilizing prospect of a ``perfect action'' for lattice
QCD.  Such an action would yield cut-off independent physical
predictions on relatively coarse-grained lattices. What they have done
is to develop a workable procedure for following the renormalization
trajectory of an asymptotically free theory from the continuum fixed
point out to relatively large values of the coupling. They have
convincingly shown that their procedure produces a ``perfect action''
for the $d=2$, $O(3)$ non-linear $\sigma$-model whose range of
interactions is short and whose structure is relatively simple.  They
are currently applying their ideas to Yang-Mills theory in four
dimensions and are finding encouraging preliminary results\cite{ahas}.
It must be said, however, that a perfect action can only be used to
answer questions which refer to modes which have not been integrated
out so that their usefulness for simulating heavy quarks may be more
limited than it is for light quarks and gluons.

\medskip
{\it Effective Lattice Field Theories: Static Quarks}

Up to now we have been considering different ways of improving the
Wilson action. Another approach is to give up trying to treat heavy
quarks as fully relativistic fermions and to expand physical
quantities in inverse powers of the mass, $m_Q$, of the heavy quark,
thereby getting rid of large $\ord{am_Q}$ discretization errors.
The zeroth order in this expansion constists in treating heavy
quarks as static, spin-1/2 color sources\cite{statapprox}. 
The corresponding action is a discretization of
\beq
S^{stat}=\int Q^\dagger D_o Q
\ ,
\eeq
where $Q$ is a two-component spinor field. The propagator for such a
quark is trivial as it is essentially a product of link variables in
the time direction. This makes static quarks very easy to simulate, at
least in principle. One can show\cite{lepagehq,statnoise}, however, that the
ratio of signal to noise for the propagator of a hadron containing a
static and a light quark falls exponentially and very rapidly in time 
so that correlators made from static propagators are usually extremely
noisy. Furthermore, because heavy quarks are static in this
approach, we cannot study transitions between heavy quarks of
different momentum as one must to describe decays such as
$\btodsl$. 

\medskip
{\it Effective Lattice Field Theories: HQET}

In this approach, one allows the heavy quark to have a non-vanishing
velocity $\s v$ and be slightly off-shell. To leading order in
heavy-quark mass, the relevant action is a discretized version of
\beq
S^{HQET}=\int Q^\dagger v_\mu D_\mu Q
\ ,
\label{shqet}
\eeq
where $v_o=\sqrt{1+\s v^{\,2}}$ and $Q$ is the same two component spinor
field as above. Obviously, when $\s v=0$, we recover $S^{stat}$.

Like the static action, this action is much easier to simulate than
the full action, because calulating the quark propagators is an
initial value problem.  The HQET moreover permits the study of
transitions between heavy quarks of different momentum which is
necessary to obtain the form factors relevant for weak $b\to c$ decays
in the heavy-quark limit. However, the low signal to noise ratio
problems encountered with static quarks are present here too.

One may also wonder whether this theory is even defined in Euclidean
spacetime as the free, momentum space propagator has a pole at 
$p_o=i\s v\cdot\s p/v_o$ which may be on the positive
or negative imaginary axis depending on the relative orientation of 
$\s v$ and $\s p$. This means that the free theory has solutions which
grow exponentially in time. One can show, however, that the resulting
divergences are regulated when these propagators are combined
with relativistic, light-quark propagators in correlation functions for
hadrons containing heavy and light quarks\cite{aglie}.

\medskip
{\it Effective Lattice Field Theories: Non-Relativistic QCD (NRQCD)}

In this approach, in addition to expanding the QCD action in inverse
powers of the heavy-quark mass, one further expands it in powers of
the heavy-quark's velocity squared, $\s v^{\,2}$, which is assumed to
be small. This double expansion leads, at order $\lqcd/m_Q$ and $\s
v^{\,2}$, to the following action\cite{hlnrqcd,cthd}
\beq
S^{NRQCD}=\int Q^\dagger \l\{D_o +\frac{\s
D^2}{2m_Q}+\frac{g_o}{2m_Q}\,\s\sigma\cdot\s B\,\r\}Q
\ ,
\label{snrqcd}
\eeq
where $\s B$ is the chromomagnetic field and $Q$ is, again, a two
component spinor field.

This action permits one to simulate the $b$ quark directly. It is also
simpler to simulate than the full theory, for the same reasons
$S^{HQET}$ is. Furthermore, the additional kinetic term in the
action guarantees that the ratio of signal to noise in correlators which
include non-relativistic propagators is much better than in correlators
which include static propagators\cite{statnoise}. 

One has to remember, however, the NRQCD is an effective theory and
that the cutoff must be kept finite. Since the expansion which leads
to the action of \eq{snrqcd} improves as $am_Q$ becomes larger,
$a^{-1}$ should not be chosen much bigger than about $2\gev$ when the
$b$-quark is being considered and should be chosen much smaller when
the $c$-quark is involved.  This means that one has to begin worrying
about discretization errors. Since these errors cannot be removed by
going to the continuum limit, they must be removed by higher dimension
operators. Whether such a program can successfully be carried out for
both spectral quantities and matrix elements has yet to be shown.
Moreover, from a more practical point of view, the amount of phase
space that can be explored in $B\to D$ decays, for instance, will be
severely limited by the fact that one has to limit the size of the
momenta given to the particles to keep momentum-dependent
discretization errors under control and for the non-relativistic
approximation to remain valid.

\subsubsection{Finite Volume Errors}

These errors are due to the fact that we approximate spacetime by a
box of finite volume. The first requirement on the size of this box
is, of course, that it be large enough to accommodate the hadron we
wish to study. If this is not the case, volume corrections which fall
off with inverse powers of the lattice's volume
appear\cite{aoki,fukugita}. For mesons composed of a heavy quark and a
light antiquark, the requirement that the box be larger than the meson
means that $aL\gg 1/\lqcd$, where $L$ is the number of lattice sites
in the space directions, since the typical size of these mesons is on
the order of $1/\lqcd$.

The second requirement, in the case of peridic boundary conditions, is
that the box be larger than the range of the strong interaction so
that the hadron cannot interact with its many
copies\cite{fukugita,luscher}.  Since this range is determined by the
mass of the lightest hadron--the long range component of the strong
interaction is governed by pion exchange and falls off exponentially
with distance as ${\rm e}^{-m_\pi\,r}$-- we must ensure that copies
are at the very least one pion compton wavelength away.  Because there
is a limit on the size of the lattices that we can work with, this
requirement is usually turned around and used to constrain the mass of
the light quarks. If we want our finite volume errors to be smaller
than about 5\%, neighboring copies must be at least three pion
wavelenghts away (${\rm e}^{-3}\simeq 5\%$). This means that the mass
of the lightest allowed quark must be such that a pseudoscalar meson
which contains this quark and its antiquark has a mass $m_P$ with
\beq
\frac{3}{m_P}\le aL-\frac{1}{\lqcd}
\ .
\label{mpimin}
\eeq
For the lattice used by the UKQCD collaboration ($a^{-1}\sim 2.7\gev$
and $L=24$) we would therefore require that the lightest pseudoscalar
meson have a mass larger than $\sim 600\mev$ ($\lqcd\simeq 250\mev$).
This constraint can, in fact, be slightly relaxed since we work in the
quenched approximation.\footnote{In the quenched approximation, a
virtual pion cannot be created out of the vacuum. Thus, the range of the
strong interaction must be determined by either gluonic modes or the
lightest hadron that can be formed from the available valence quarks.}

Before leaving the subject, I would like to point out that our
collaboration has recently uncovered an unexpected finite volume
volume effect\cite{juan} in trying to implement a new idea for
obtaining the slope of the Isgur-Wise function at zero
recoil\cite{slope}. This effect is purely kinematical: it is solely
due to the fact that the momentum of a quantum particle in a box is
quantized.  Because of its purely kinematical nature, we were able to
correct it analytically and extract sensible results even though the
distortions this effect gave rise to were as large large as 30\%. For
more details, please see Ref.~\cite{juan}.

\subsubsection{Quenching}
As mentioned earlier, most calculations of weak matrix elements on the
lattice are performed in the quenched approximation. This means that
the fermion determinant, ${\rm det}\l(\Ds+M\r)$, is set to 1 when
generating the gluon configurations according to the distribution
given in \eq{dist}. Physically, this approximation corresponds to
neglecting the effect of quark loops. The errors this approximation
induces are difficult to estimate and will depend on the quantity studied.
Experience shows, however, that the
results of quenched calculations generally agree with experiment at a
level ranging from appoximately 0 to 20\%. It should be noted, also,
that quite a bit of progress has been made in understanding the chiral
behavior of quenched QCD by designing and exploring a quenched version
of the traditional chiral lagrangian. (Please see M.~Golterman's
lectures in the present volume for a very clear introduction to
the subject.)

The reason why so many lattice groups work in the quenched
approximation is because, with present day algorithms, the CPU time
required for a quenched simulation is roughly proportional to the number of
points on the lattice while for an unquenched simulation, this time is
roughly proportional to the number of points to the power 2.5.

\subsubsection{Matching}
On the lattice, the local vector and axial-vector currents 
$\bar q'\g^\mu q$ and $\bar q'\g^\mu\g^5 q$, where $q$ and $q'$ are
quark fields, are not symmetry
currents and can therefore develop a dependence on the cutoff through
renormalization.
\footnote{For the decay $\bksg$, the operator whose matrix element
one has to consider is already scale-dependent in the continuum
because it is an effective local operator which results from
integrating out the $W$.}  To cancel this dependence
and obtain physical results, one has to multiply the corresponding
lattice results by renormalization constants which exactly compensate
the scale-dependence of the currents.
\footnote{These two currents do not mix with other operators on the lattice
and are therefore multiplicatively renormalized.} The renormalization
constant for the vector current is known as $Z_V$ and that of the
axial-vector current as $Z_A$. Since these constants incorporate
physics that lies above the cutoff, they can be calculated
perturbatively as long, of course, as this cutoff is much larger than
$\L_{QCD}$. Thanks to the techniques developed recently for improving
the notoriously poor convergence of bare lattice perturbation
theory\cite{lepmac}, these perturbative determinations are becoming
much more reliable. One is not restricted to use perturbation theory,
however. One can determine $Z_V$ and $Z_A$ non-perturbatively by
tuning these constants so that hadronic matrix elements of the
corresponding currents satisfy lattice Ward identities\cite{ward} or
even some other normalization condition (see for example Ref.
\cite{iwprd}). For operators which are not symmetry currents in the
continuum Ward identities are not an option and the use of a
normalization condition on hadronic matrix elements represents a loss
of predictive power. It is therefore preferable to normalize such
operators using matrix elements of quarks and gluons. Techniques for
performing such non-perturbative renormalizations of arbitrary lattice
operators are currently being developed by G.~Martinelli et
al.\cite{nonpert}.  

In any case, whether perturbative or non-perturbative, this
matching of lattice and continuum operators does introduce
supplementary uncertainties whose size is determined by the accuracy
with which the matching coefficients are calculated.



\subsection{The UKQCD Lattice}
\label{latdet}
For completeness, I now briefly detail the parameters used for the
lattice calculations whose results I present in the following.  These
calculations are performed on a $24^3\times 48$ lattice at an inverse
coupling $\beta=6.2$ which corresponds to an inverse lattice spacing
$a^{-1}\simeq 2.7\gev$ as shown in \tab{scale}. We work in the
quenched approximation and use the ``clover'' action of \eq{clover} to
describe quarks. We further rotate quark fields according to \eq{rot}
to obtain improved operators. Our leading discretization errors are
thus of order $\ord{\as a,a^2}$.

To keep
volume errors under control, we perform our calculations with light
quarks whose masses are in the range of the strange-quark mass (see
\tab{light}). To obtain the corresponding quantities for
up and down quarks, we extrapolate our results in light quark
mass to the chiral limit (i.e. the limit of vanishing light-quark
mass). Results for strange light quarks are obtained by interpolation.
\begin{table}
\centering
\begin{tabular} {|c|c|}\hline
$\k_q$  & $m_P$ ($\gev$)\\
\hline
0.14144 & $\sim$ 0.80\\
0.14226 & $\sim$ 0.58\\
0.14262 & $\sim$ 0.45\\
\hline
\end{tabular}
\begin{center}
\begin{minipage}[t]{\captionwidth}
\caption{$\k_q$ is the light quark's hopping paramter and 
$m_P$ is the mass of the pseudoscalar
meson obtained by combining this quark with its antiquark
\protect{\cite{strange}}. Here, $m_\rho$ was used to set the
scale. The errors on $m_P$ are dominated by the uncertainty
in the scale and are on the order 10 to 15\%.
\label{light}}
\end{minipage}
\end{center}
\end{table}

To keep discretization errors under control, we limit our heavy
quarks to have masses about $m_c$ (see \tab{heavy}).
To obtains results which are relevant for situations where the
heavy quark is a $b$, we extrapolate our results in heavy-quark
mass to $m_b$ using HQET as a guide.
\begin{table}
\centering
\begin{tabular} {|c|c|}\hline
$\k_Q$  & $m_Q$ ($\gev$)\\
\hline
0.121 & 1.90\\
0.125 & 1.64\\
0.129 & 1.36\\
0.133 & 1.06\\
\hline
\end{tabular}
\begin{center}
\begin{minipage}[t]{\captionwidth}
\caption{$\k_Q$ is the heavy quark's hopping paramter and 
$m_Q$ is it's physical mass. This mass is obtained by subtracting the
$500\mev$ of energy carried by the light degrees of
freedom\protect{\cite{bible}} from the spin averaged mass,
$(3M_V+M_P)/4$, of a vector ($V$) and apseudoscalar meson ($P$)
obtained by combining the heavy quark with a massless light quark (see
Ref.\protect{\cite{iwprd}} for details). Here again, $m_\rho$ was used
to set the scale. The errors on $m_Q$ are dominated by the uncertainty
in the scale and are on the order 10 to 15\%.
\label{heavy}}
\end{minipage}
\end{center}
\end{table}

Finally, a pointlike operator such as $\bar Q(x)\Gamma q(x)$, where
$\Gamma$ is some combination of Dirac matrices, $Q$ is a heavy quark
and $q$ a light one, usually has very poor overlap with the ground
state of the corresponding meson since the latter has spatial extent.
Thus, to improve the overlap it is natural to use extended
operators. This usually involves ``smearing'' the heavy-quark field
over a volume which is determined by a kernel
$K(x,x^\prime)$:
\beq
Q^S(\s x,t)\equiv \sum_{\s x^\prime}K(x,x^\prime)Q(\s x^\prime,t),
\label{eq:kernel}\eeq
where $Q^S(\s x,t)$ is the smeared field. The results presented below were
obtained using a gauge-invariant
kernel which has an $rms$ radius of 5.2 lattice units (for
details, see ref.~\cite{smearing}).




\section{Leptonic Decays
\protect{\footnote{The UKQCD results presented in this section have been
published in Ref.\protect{\cite{qhldc}}.}}}
\label{fb}

\subsection{Leptonic Decays of Pseudoscalar Mesons}
\label{ldopm}
In the present section, I consider the leptonic decays of a pseudoscalar
meson, $P$, containing a heavy quark $Q$ and a light antiquark $\bar
q$. In the Standard Model these decays are mediated by a $W$ boson. As
mentioned in the Introduction, the coupling of the quarks to this
boson is strongly modified by non-perturbative QCD dynamics. This
modification of the quark-boson vertex is parametrized by a decay
constant, $f_P$, which we calculate numerically using the lattice
formulation of QCD.

Because the amplitude for these leptonic decays is proportional to the
CKM matrix element $V_{q\,Q}$ ($V_{ub}$ for the leptonic decay of a
$B^-$ meson), one might think that a study of these decays could
lead to a determination of $V_{q\,Q}$. However, one can
show that the branching ratio for these decays is suppressed by five
powers of the heavy-meson's mass in the limit that this mass becomes
very large\cite{bible}
\beq
{\cal B}\l(P\longrightarrow\ell\bar\nu_\ell\r)\toas{M_P\to\infty}\sim
\frac{m_\ell^2}{M_P^5}\,|V_{q\,Q}|^2\ f_P^2
\ ,
\eeq
and is therefore very small.
For $P=B^-$ and $\ell=\tau^-$, and for reasonable values of $f_B$
($\sim 160\mev$) and $V_{ub}$ ($\sim 0.005$), the branching ratio is
on the order of $10^{-4}$. Thus it may be a while before these
decays provide a precise determination of $V_{ub}$.  Nevertheless, a
precise knowledge of the corresponding decay constants is important.
These constants are required for describing $\bar B-B$ mixing
\footnote{The important non-perturbative quantity here is, in fact,
$f_B\,B_B$ where $B_B$ is known as the ``bag'' parameter. The latter
is actually currently being calculated by the UKQCD Collaboration in
the limit of a static $b$-quark.} as well as for describing
non-leptonic weak decays in factorization
schemes\cite{ref108109}. Furthermore, a detailed study of how $f_P$
depends on the mass of the heavy quark, $Q$, provides an important
tool for gauging the range of applicability of heavy-quark symmetry.

To describe these leptonic decays, we must calculate the non-perturbative
matrix element
\beq
\la 0|\,\bar q\g_\mu\g_5 Q(0)\,|P(\s p)\ra=ip_\mu\,f_P
\ .
\eeq
As sketched in \sec{what}, we obtain this matrix element from the 
lattice 2-point function
\beq
\sum_{\s x}\la\bar Q^R\g_\mu\g_5q^R(x)\,\bar q^R\g_5 Q_s^R(0)\ra
\ ,
\eeq
where the subscript $s$ indicates that the corresponding field has
been smeared as in \eq{eq:kernel} while the superscript $R$ indicates
that the field has been rotated according to \eq{rot}. We evaluate
this correlator for the four heavy-quark flavors described in
\tab{heavy} and for the three light-quark flavors described in
\tab{light}. Since the heavy-quark flavors we consider all have masses
in the range of the charm, we do not obtain $f_B$ directly. In order
to get $f_B$, we must first understand how the decay constant $f_P$
varies with the mass of the heavy quark. 

HQET predicts that the quantity
\beq
{\hat\Phi}(M_P)\equiv\l(\frac{\as(M_P)}{\as(m_B)}\r)^{2/\beta_o}Z_A^{-1}
\,f_P\sqrt{M_P}
\label{phimp}
\eeq
is independent of $M_P$ up to corrections proportional to inverse
powers of $M_P$ and up to higher-order radiative corrections:
\beq
{\hat\Phi}(M_P)\toas{M_P\to\infty} {\rm cst}+
\ord{\frac{\lqcd}{M_P},\,\a_s(M_P)}
\label{phimplim}
\ .\eeq

\begin{figure}[t]
\begin{center}
\leavevmode
\BorderBox{2pt}{%
\InsertFigure[140 480 480 800]{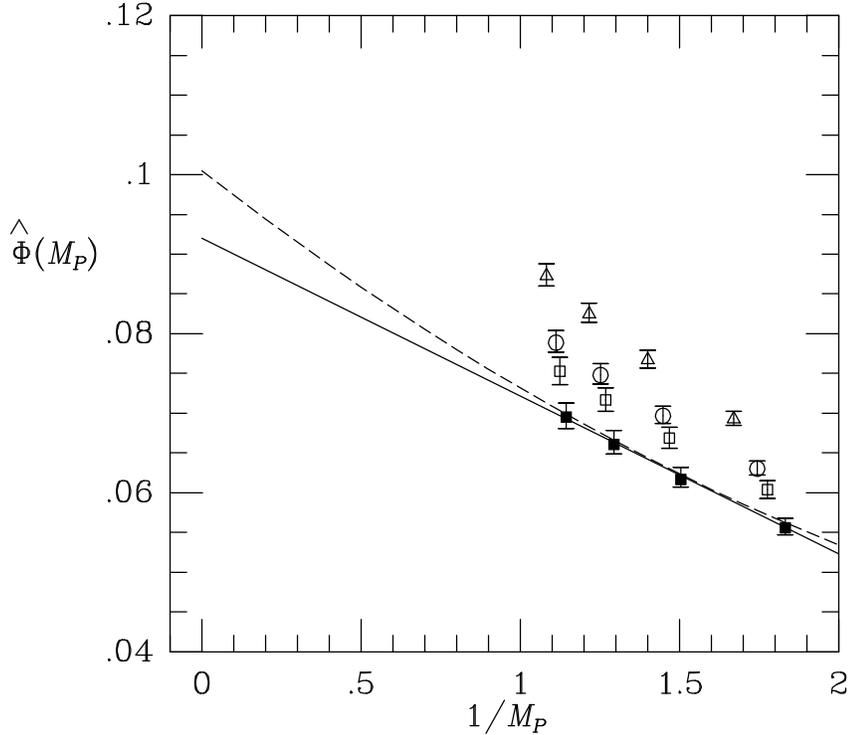}{340bp}{320bp}%
}
\begin{minipage}[t]{\captionwidth}
\caption{UKQCD data for $\protect\hat{\Phi}(M_P)$
plotted against inverse meson mass. The open symbols correspond to
the decay constants obtained for every possible combination of our
four heavy quarks mass and our three 
light antiquarks. The full symbols, on the other hand, are obtained
by extrapolating the light-antiquark mass to zero at fixed heavy-quark
mass. The solid line represents the linear
fit to the chirally-extrapolated points using the three heaviest meson
masses, whereas the dashed curve results from a quadratic fit to all
four.\label{phimpvsmp}}
\end{minipage}
\end{center}
\end{figure}

To evaluate ${\hat\Phi}(M_P)$ we take $\beta_o=11-2/3n_f$ with $n_f=0$ and
use for $\as$ the one-loop expression with $\lqcd=200\mev$. We
plot our results for ${\hat\Phi}(M_P)$ in \fig{phimpvsmp} as a function of
$1/M_P$. 
%

In the heavy-quark limit, ${\hat\Phi}(M_P)$ is a horizontal line as
\eq{phimplim} indicates. The fact that ${\hat\Phi}(M_P)$ increases quite
sharply with $M_P$ in \fig{phimpvsmp} is a clear indication that 
corrections to the heavy-quark limit are quite large. To quantify
these corrections we fit ${\hat\Phi}(M_P)$ to a linear and a quadratic
function of $1/M_P$.  These fits indicate that the $1/m_Q$-corrections
to $f_P$ are on the order of 10-15\% at the scale of the $B$ and 27-40\% at
the scale of the $D$. This larger than one would expect
on the naive grounds that these correction should be of order
$\lqcd/m_b\simeq 0.25/4.8\simeq 5\%$ for $f_B$ and of order
$\lqcd/m_c\simeq 0.25/1.45\simeq 17\%$ for $f_D$, a phenomenon also
observed in sumrule calculations where these corrections range
from 13\% to 22\% at $m_B$ and from 37\% to 64\% at 
$m_D$\cite{bib104138139}.

Having determined how $f_P$ depends on $M_P$, we interpolate $f_P$ in
$M_P$ to the $D$ and extrapolate to the $B$. We present our results
for the corresponding decay constants $f_D$ and $f_B$ in \tab{fbres}.
We also provide results for $f_{D_s}$ and $f_{B_s}$ obtained in the
same way as $f_D$ and $f_B$ but with the light quark fixed to be the
strange.  We further give, for comparison, the results obtained by
other lattice groups as well as experimental measurements when
available.
\begin{table}
\centering
\begin{tabular} {|c|c|c|c|c|c|}\hline
Ref.& Action & $\beta$&$f_D(\mev)$&$f_{D_s}(\mev)$ & $f_{D_s}/f_D$ \\ \hline
UKQCD\cite{qhldc} & Clover & 6.2 & $185\er{4}{3}\err{42}{7}$ & $212\er{4}{4}
\err{46}{7}$ &$1.18\er{2}{2}$\\ \hline
APE\cite{ape} & Clover & 6.0 & $218\pm 9$ & $240 \pm 9$ & \\ \hline
BLS\cite{bernard} & Wilson &6.3&$208(9)\pm 35\pm 12$&$230(7)\pm 30 \pm 18$
& $1.11(2)\pm.04\pm.02$\\ \hline
ELC\cite{orsw} & Wilson & 6.4 & $210 \pm 15$ & $227 \pm 15$ &\\ \hline
ARGUS\cite{argusfd}&&&-&$267 \pm 28$ & \\ \hline
CLEO\cite{cleofd}&&&-&$344 \pm 37\pm 52 \pm 42$ &\\ \hline
WA75\cite{wa75}&&&-&$232 \pm 45 \pm 20 \pm 48$ &\\ \hline
\hline
Ref.& Action & $\beta$ & $f_B(\mev)$ & $f_{B_s}$ & $f_{B_s}/f_B$ \\ \hline
UKQCD\cite{qhldc} & Clover & 6.2 & 
$160\er{6}{6}\err{53}{19}$ & $194\er{6}{5}\err{62}{9}$ & $1.22\er{4}{3}$
\\ \hline
APE\cite{ape} & Clover & 6.0 & $197\pm 18$ & & \\ \hline
BLS\cite{bernard} & Wilson &6.3 &
 $187(10)\pm34\pm15$ & $207(9)\pm34\pm22$ &$1.11(2)\pm.04\pm.02$\\ \hline
ELC\cite{orsw} & Wilson &6.4 &$205\pm40$ & & $1.06\pm.04$\\ \hline
HEMCGC\cite{hem}& W+S& 5.6 & $200\pm48$ & &\\ \hline
\end{tabular}
\begin{center}
\begin{minipage}[t]{\captionwidth}
\caption{Heavy-light decay constants obtained with propagating, relativistic
heavy quarks. The normlization used here is the one for which
$f_\pi=132\protect{\mev}$.
The HEMCGC calculation is unquenched and was performed
with Wilson valence quarks and Staggered sea quarks.
\label{fbres}}
\end{minipage}
\end{center}
\end{table}

For completeness, finally, we provide in \tab{fbres} 
the ratios of decays constants $f_{D_s}/f_D$ and $f_{B_s}/f_B$
in which many systematic errors cancel.

I should mention that there is a very vast literature on the subject
of computing $f_B$ in the static approximation of Lattice QCD (see,
for instance, the reviews of Ref.\cite{brev}) and
that a few groups are beginning to compute this same quantity using
the NRQCD action of \eq{snrqcd}\cite{cthd,nrqcdfb}.

\subsection{Leptonic Decays of Vector Mesons}
We now consider the leptonic decays of a vector meson, $V$, composed of
a heavy quark, $Q$, and a light antiquark, $\bar q$. These decays
are not relevant phenomenologically as a $B^*$ or $D^*$ meson
will decay electromagnetically or through the strong interaction
long before it decays weakly. They are interesting, however,
because they provide a means of testing heavy-quark symmetry.
HQET
predicts that the quantity
\beq
\tilde U(M)=\frac{f_Vf_P}{M}\,\frac{1}{1+\frac{2}{3\pi}\as(M)}
\ ,
\label{utilde}
\eeq
is equal to 1 up to corrections proportional
to inverse powers of the mass $M$ and higher order radiative corrections:
\beq
\tilde U(M)\toas{M\to\infty} 1+\ord{\frac{\lqcd}{M},\,\a_s^2(M)}
\ .
\label{ulim}
\eeq
Here, $M=(3M_V+M_P)/4$ and $f_V$ is the decay constant given by
\beq
\la0|\bar q\g^\mu Q(0)|V,\e\ra=\e^\mu\frac{M_V^2}{f_V}
\ ,
\eeq

\begin{figure}[t]
\begin{center}
\leavevmode
\BorderBox{2pt}{%
\InsertFigure[140 480 480 800]{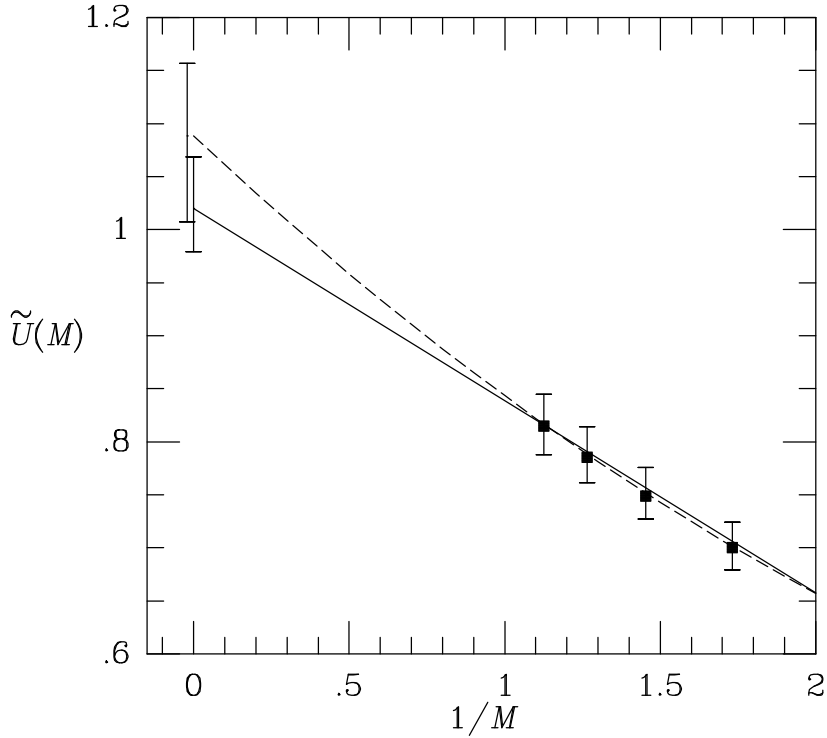}{340bp}{320bp}%
}
\begin{minipage}[t]{\captionwidth}
\caption{The quantity $\protect\tilde{U}(M)$ plotted against the inverse
spin-averaged mass. 
Linear and quadratic fits are represented by the solid and dashed
curves, respectively. Also shown are the statistical errors of the
extrapolation to the infinite mass limit. In this plot, the
light antiquark is masseless.\label{fig:utilde}}
\end{minipage}
\end{center}
\end{figure}

In \fig{fig:utilde}, our results for the quantity $\tilde U$ are
plotted as a function of $1/M$. If heavy-quark symmetry were respected
in the charm sector, we would expect all four points to lie on the
horizontal line $\tilde U=1$. The fact that $\tilde U(M)$ is not
horizontal in \fig{fig:utilde} is again an indication that heavy-quark
symmetry is quite badly broken. To quantify this statement, we peform,
as we did for ${\hat\Phi}(M_P)$, a linear and a quadratic fit in $1/M$ to
our data.  We find that corrections to the heavy-quark limit are again
on the order of 10\% at the scale of the $B$ and on the order of 30\%
at the scale of the $D$.

The fact that $\tilde U$ extrapolates close to 1 when $1/M\to 0$--i.e.
the value it is supposed to take in the heavy-quark limit-- provides
support for our description of power corrections. It also gives us
confidence that discretization as well as other systematic errors are
under control.


\section{Semi-Leptonic Decays}
\label{semi-lep}

\begin{figure}[t]
\begin{center}
\leavevmode
\centerline{
\ewxydiag{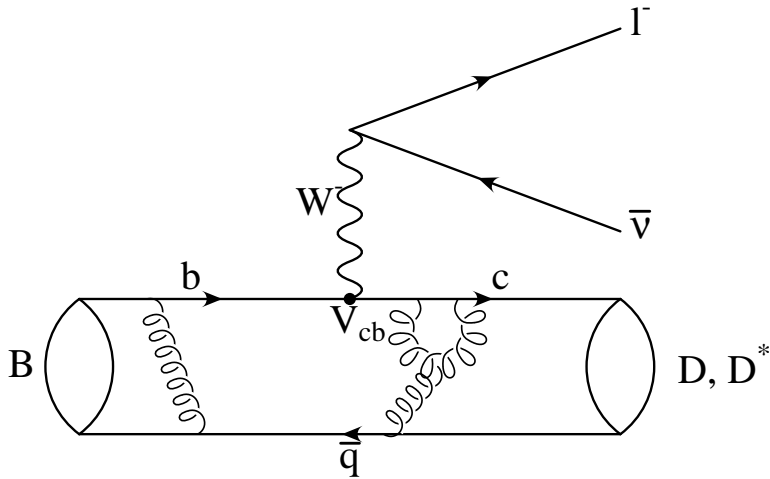}{350pt}
}
\begin{minipage}[t]{\captionwidth}
\caption{Semi-leptonic decays of a 
$B$ meson. Here, $l^-$ is a lepton and $\bar\nu$ is its
anti-neutrino.  
\label{feyn_btod}}
\end{minipage}
\end{center}
\end{figure}

Let us now turn to the subject of semi-leptonic decays of $B$ mesons
into $D$ or $D^*$ mesons. These decays are depicted in 
\fig{feyn_btod}.
As was the case for leptonic decays, the fact
that the $b$ and the $c$ are confined within hadrons severely modifies
their coupling to the $W$. 

These semi-leptonic decays are interesting theoretically.  They enable
one to determine the now famous Isgur-Wise function and to test the
range of validity of heavy-quark symmetry.  But they are also
interesting phenomenologically because they can be used to measure $V_{cb}$.
The decay $\btodsl$ is, in fact, very well suited for this task
because its rate, aside from being rather large, is free of
$\ord{\lqcd/m_c}$-corrections to the heavy-quark symmetry prediction
at maximum four-momentum transfer (Luke's theorem\cite{luke}):
\beq
\frac{1}{\sqrt{\w^2-1}}\frac{d\Gamma(\btodsl)}{d\w}
\toas{w\to 1}
\frac{G_F^2|V_{cb}|^2}{4\pi^3}(m_B-m_{D^*})^2m_{D^*}^3\eta_A^2\,\l(1+\ord{
\l(\frac{\lqcd}{m_c}\r)^2}\r)
\ ,
\label{ratelim}
\eeq
where $\w=v_B\cdot v_{D^*}$ is the recoil, $v_{B(D^*)}$ are the
four-velocities of the mesons and $\eta_A$ are perturbative radiative
corrections that will be described shortly. Thus, a measurement of the
decay rate very close to the zero-recoil point yields a
model-independent determination of $V_{cb}$\cite{bible}
up to small, non-perturbative $\ord{\l(\frac{\lqcd}{m_c}\r)^2}$
corrections.  In practice, what one does is to measure the
differential decay rate over the whole kinematical range and
extrapolate it to $\w=1$. To be reliable, this extrapolation requires
a theoretical guide, especially since the different experiments
(ARGUS, CLEO and ALEPH) have been finding decay rates with rather
different $\w$-dependences (see Figs. \ref{vcbcleo}, \ref{vcbaleph}
and \ref{vcbargus}). This in turn requires a knowledge of
the rate's dependence on $\w$ which is governed by the
non-perturbative QCD dynamics which binds the quarks into mesons. It
is here that our lattice calculation enters.

$\btodl$ decays are slightly less well suited for obtaining $V_{cb}$
as they are helicity suppressed,
\footnote{The rate is down by a factor of $\w^2-1$ compared to that
for $\btodsl$ decays.} which makes them more difficult to measure
close to $\w=1$. Moreover, their rate is not protected by Luke's
theorem and should suffer larger $\ord{\lqcd/m_c}$ corrections close
to zero recoil. Neither of these problems are insurmountable, however,
since the rate is far from negligible away from $\w=1$ and the
$\ord{\lqcd/m_c}$ corrections turn out to be parametrically
suppressed\cite{bible}. So in the future, when better data is available,
this channel should also provide a good means for determining
$V_{cb}$.

\bigskip
The matrix element required to describe $\btodl$ and $\btodsl$ decays are
\beq
\frac{\la D(v')|\bar c\g^\mu b|B(v)\ra}{\sqrt{m_Bm_D}}=(v+v')^\mu\,h^+(\w)
+(v-v')^\mu\,h^-(\w)
\ ,
\label{bd}
\eeq
\beqa
&&\frac{\la D^*(v',\e)|\bar c\g^\mu b|B(v)\ra}{\sqrt{m_Bm_{D^*}}}=
i\e^{\mu\nu\a\b}\e^*_\nu v'_\a v_\b\,h_V(\w)\nonumber\\
\mbox{and}&&\\
&&\frac{\la D^*(v',\e)|\bar c\g^\mu\g^5 b|B(v)\ra}{\sqrt{m_Bm_{D^*}}}=
(\w+1)\e^{*\mu}\,h_{A_1}(\w)-\e^*\cdot v\l(v^\mu\,h_{A_2}+v'^\mu\,h_{A_3}\r)
\nonumber\ ,
\label{bds}
\eeqa
where $\w=v\cdot v'$ and $\e^\mu$ is the polarization vector of the
$D^*$.

In the limit that $m_c$ and $m_b$ are infinite, 
heavy-quark symmetry reduces
the six form factors of \eqs{bd}{bds} to a single universal
function of the recoil, $\xi(\w)$, known as the Isgur-Wise function:
\beqa
&& h_+(\w) =h_{A_1}(\w)=h_{A_3}(\w)=h_V(\w)=\xi(\w)\nonumber\\
\mbox{and}&& \\
&& h_-(\w)= h_{A_2}(\w)\equiv 0
\nonumber\ . 
\label{ffeq}
\eeqa
This symmetry further requires that the Isgur-Wise function be
normalized to one at zero recoil\cite{iw}:
\beq
\xi(1)=1
\ .
\eeq

Of course, in nature, the bottom and the charm are not infinitely
massive and there are corrections to the simple relations of
\eq{ffeq} as there were to the scaling relation of \eqs{phimplim}{ulim}. 
These corrections can be parametrized in terms of
twelve functions $\b_i(\w)$ and $\g_i(\w)$ with $i=+,\ -,\ V,\ A_1,\
A_2,\ A_3$ such that
\beq
h_i(\w)=\l(\a_i+\b_i(\w)+\g_i(\w)\r)\ \xi(\w)
\ ,
\eeq
where $\a_+=\a_V=\a_{A_1}=\a_{A_3}=1$ and $\a_-=\a_2=0$. The functions
$\b_i$ correspond to the exchange of hard gluons across the vector and
axial-vector vertices. These corrections are perturbative and in the
sequel we use the results obtained by M.~Neubert in Ref.\cite{short}.
In this reference, Neubert performs a full one-loop matching of HQET
to QCD and runs the results at two loops.

The second set of corrections, the $\g_i$'s, are proportional to
inverse powers of the heavy-quark masses $m_c$ and $m_b$. They
correspond to matrix elements of higher-dimension operators in HQET
and are therefore non-perturbative and almost as difficult to evaluate
as full QCD matrix elements. If most of these corrections cannot be
neglected, heavy-quark symmetry obviously loses much of its 
predictive power. In
most cases, heavy-quark symmetry is not a precision tool
since leading power corrections are expected to be on
the order of $\lqcd/m_c\simeq 20\%$.  In a few instances, however,
heavy-quark symmetry is more constraining as we saw at the beginning of the
present section when discussing the extraction of the CKM matrix
element $V_{cb}$ and Luke's theorem. 

At the level of form factors, Luke's theorem
guarantees that $h_+$ and $h_{A_1}$ do not
suffer $\ord{\lqcd/m_{c,b}}$ corrections at zero recoil. Thus
\beq
\g_+(1),\ \g_{A_1}(1)=\ord{\l(\frac{\lqcd}{m_{c,b}}\r)^2}
\label{lukethm}
\eeq
which means that the leading non-perturbative corrections on these
form factors ought to be on the order of 4\%--i.e., very small. 

It is important to emphasize, at this point, that to 
claim that the Isgur-Wise function, $\xi(\w)$, has been obtained from
any one of the form factors, $h_i(\w)$, one has to convincingly
argue that one controls the non-perturbative power corrections, $\g_i(\w)$.


\subsection{The Form-Factor $h^+(\w)$
\protect{\footnote{Most of the UKQCD results presented in this section
will appear in Ref.\protect{\cite{iwprd}}.}}}
\label{hpanal}

The form factor $h^+$ dominates the rate for $\btodl$ decays.  
The matrix element of \eq{bd} which $h^+$ parametrizes is obtained from
the 3-point function
\beq
\sum_{\s x,\s y}\,e^{-i\s q\cdot\s x}e^{-i\s p'\cdot\s y}\,\la
\bar q^R\g_5c^R_s(t_f,\s y)\;\bar c^R\gm b^R(t,\s x)\;
\bar b^R_s\g_5 q^R(0,\s 0)\ra
\ ,
\label{thpthp}
\eeq
as sketched in \sec{what}.  (See \eqs{eq:kernel}{rot} for the meaning
of subsrcript $s$ and the superscript $R$.)

Since the heavy-quark flavors we consider all have masses
in the range of the charm, we do not directly obtain the form factor
$h^+(\w)$ relevant for physical $\btodl$ decays.  In order to get this
physical form factor, we must extrapolate the mass of the initial
heavy quark to $m_b$ and interporlate the mass of the final heavy
quark to $m_c$. Thus, we must understand how $h^+$ depends on the
mass of the initial and final heavy quarks. Since we already
understand the dependence on mass which comes from radiative corrections,
we study the quantity
\beq
\frac{h^+(\w)}{1+\b^+(\w)}\simeq \l(1+\g^+(\w)\r)\,\xi(\w)
\ ,
\label{hpoverrad}
\eeq
where these corrections are subtracted. In this equation, equality
holds only to leading order in power and radiative corrections.

\begin{figure}[t]
\begin{center}
\leavevmode
\centerline{
\ewxy{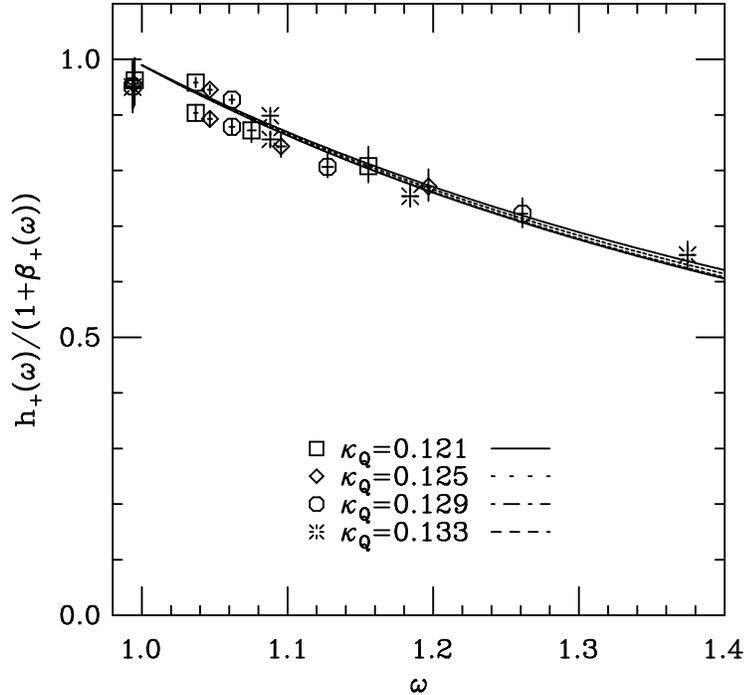}{80mm}
}
\begin{minipage}[t]{\captionwidth}
\caption{The ratio $h^+(\w)(1+\b^+(\w))$ is plotted as a function
of $\w$ for four different degenerate transitions corresponding to
four values of the heavy-quark mass. The data corresponding to each
transition has it's own symbol as detailed on the plot.  The
different curves result from individually fitting each set of data to
the parametrization $s\xi_{NR}$ (see text). The hopping parameter of
the light, spectator antiquark is $\k_q=0.14144$.
\label{hp_degen}}
\end{minipage}
\end{center}
\end{figure}

In \fig{hp_degen} we plot the ratio $h^+/(1+\b^+)$
for four degenerate transitions. 
A degenerate
transition is one in which the masses of the initial and final heavy
quarks are equal. We consider these transitions first, because they
are more constrained theoretically\cite{iwprd}.
\footnote{Because of electromagnetic current conservation, degenerate
transitions have $\g^+(1)=0$ to all orders in $1/m_{b,c}$ and
$h^-(\w)\equiv 0$ for all $\w$.}
The fact that all four sets
of points, which correspond to heavy-quark masses ranging from
$1.1\gev$ to $1.9\gev$, appear to lie very much on the same curve
is a first indication that the heavy-quark-mass dependence of $h^+/(1+\b^+)$ 
is very small. That all four sets of data lie on the same curve 
can actually be shown by fitting
each set individually to a standard parametrization for the 
Isgur-Wise function
\footnote{If the heavy-quark mass dependence of $h^+/(1+(1+\b^+))$
is negligible this quantity is just the Isgur-Wise function
$\xi$.} and showing that the parameters of the resulting curves are
equal within statistical errors. The parametrization we use
is\cite{neurieck}
\beq
\xi_{NR}(\w)=\frac{2}{\w+1}\;{\rm exp}\l[-(2\rho^2-1)\,\frac{\w-1}{\w+1}\r]
\ ,
\label{xinr}
\eeq
where $\rho^2$ is the negative of the slope of the function $\xi_{NR}$
at $\w=1$ and is the parameter of this fitting function.  We acutally
fit the data to $s\,\xi_{NR}$ where $s$ is a parameter which allows us
to test how well we have normalized our data. If our data is well
normalized and $\xi_{NR}$ is a valid parametrization for the data,
then $s$ ought to be equal to 1. For all four sets of data, we find
$s=0.99(1)$ and $\rho^2=1.4$ where the error on $s$ is statistical and
the statistical errors on $\rho^2$ are the order of $0.2$ but differ
slighly from data set to data set. Thus, our data appears to be 
correctly normalized and independent of heavy-quark mass.

\begin{figure}[t]
\begin{center}
\leavevmode
\centerline{
\ewxy{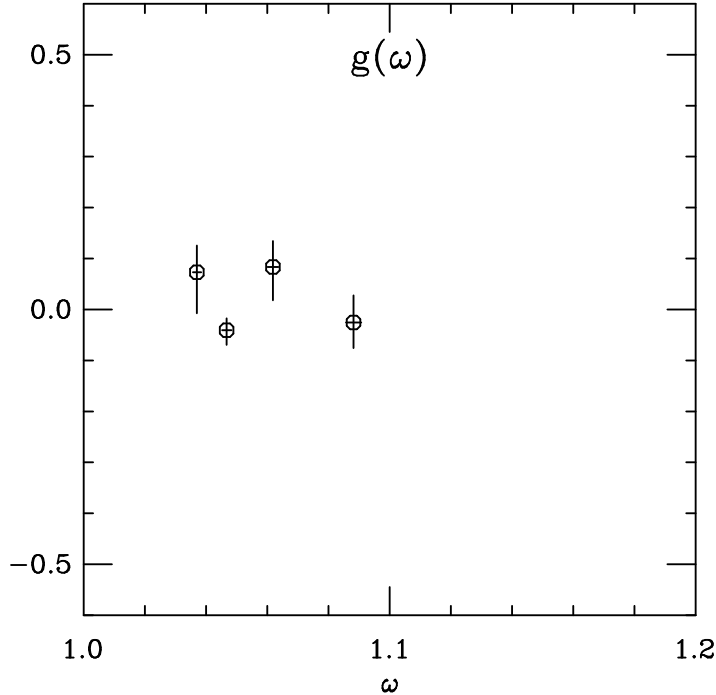}{80mm}
}
\begin{minipage}[t]{\captionwidth}
\caption{The subleading form factor, $g(\w)$, is plotted versus
$\w$. The hopping parameter of the light, spectator antiquark
is $\k_q=0.14144$.
\label{g_p_vs_w}}
\end{minipage}
\end{center}
\end{figure}

This mass dependence can in fact be quantified by extracting the
power corrections, $\g^+$, from our data. If radiative corrections
are neglected, these power corrections can be
parametrized in terms of a single universal function, $g(\w)$, such
that
\beq
\g^+(\w)=g(\w)\,\frac{\bar\L}{2}\,\l(\frac{1}{m_b}+\frac{1}{m_c}\r)
+\ord{\l(\frac{\bar\L}{2m_{c,b}}\r)^2}
\ ,
\label{gdef}
\eeq
where $\bar\L$ is the energy carried by the light degrees of freedom and
is approximately equal to $500\mev$ when the light quark $q$ is an up
or down quark\cite{bible}. Furthermore, according to 
Luke's theorem, 
\beq
g(1)=0
\ .
\eeq
The form factor, $g(\w)$, can be extracted from our results for $h^+$
by taking ratios of the quantity $h^+/(1+\b^+)$ at fixed $\w$ but
different values of the initial or final heavy quark mass (see
\eq{hpoverrad}). Because
lattice momenta are quantized, there are, in our data, only four
values of $\w$ for which this can be done. These four point are
plotted in \fig{g_p_vs_w}.
This figure indicates that $g(\w)$ is consistent with 0 and in any
case is less than one or two times $10^{-1}$ in the range of recoils
that we can explore. Since $(\bar\L/2)(1/m_b+1/m_c)$ is about 0.4 for
the quarks we simulate power corrections to $h^+$ appear to be indeed
very small.  When combined with the information obtained earlier for
degenerate transitions which covered a much larger range of recoils,
this result clearly confirms that $h^+/(1+\b^+)$ depends very little
on heavy-quark mass. This weak dependence on heavy-quark mass,
however, can be interpreted in many ways. A first explanation is that
power corrections to $h^+$, as well as discretization errors
proportional to the masses of the heavy quarks, are genuinely small. A
second possible interpretation is that discretization errors are small
but that the $\ord{\bar\L/2m_{b,c}}$-correction happens to cancel
against the higher order power corrections in the range of heavy-quark
mass that we are investigating--i.e. that in this range of masses the
heavy quark regime has not yet been reached. A third is that the power
corrections and discretization errors have opposite signs and cancel.
Since the cancellations of options two and three would have to take
place over a rather large range of recoils and a rather large range of
heavy-quark masses, they seem quite unlikely. So in what follows, we
will adopt the interpretation that power corrections to $h^+$
are genuinely small even for heavy quark with mass close to that of
the charm.
\footnote{These arguments are fleshed out in Ref.\cite{iwprd}. It
must also be said that the way in which we normalize our data means
that we measure $\g^+(\w)-\g^+(1)$ and not $\g^+(\w)$.
Thus, we are not sensitive to those 
power corrections which depend very weekly on $\w$. If these
corrections happen to be large, our conclusions about the size of
power corrections are invalid.  However, because our normalization
procedure subtracts some of the higher-order power corrections, our
determination of the Isgur-Wise function as well as our determination
of the leading $\ord{\bar\L/2m_{b,c}}$ corrections should be all the
more accurate. Again see
Ref.\cite{iwprd} for details.} 

\begin{figure}[t]
\begin{center}
\leavevmode
\centerline{
\ewxy{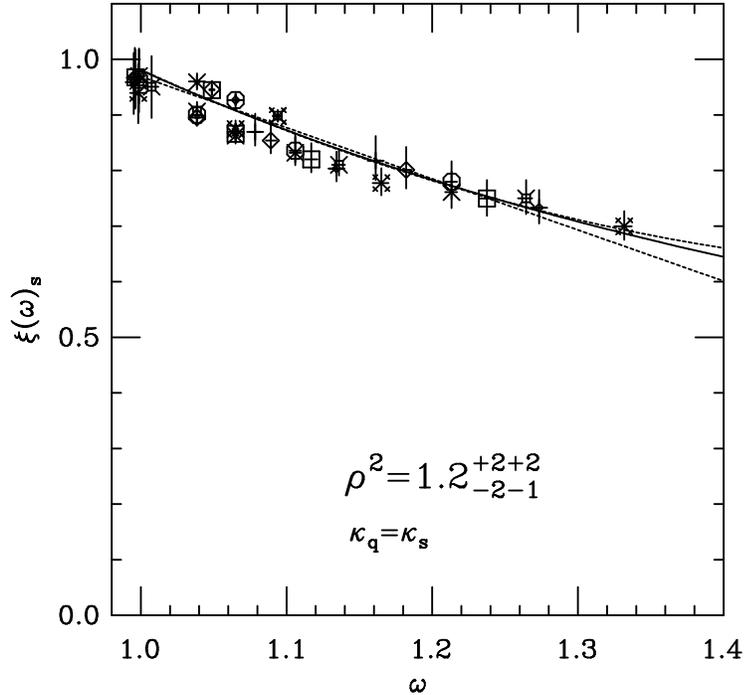}{80mm}
}
\begin{minipage}[t]{\captionwidth}
\caption{The ratio $h^+/(1+\b^+)$ is plotted as a function of $\w$. 
The different symbols correspond to different initial and/or final
heavy-quark masses. All symbols, however, correspond to situations
where the light antiquark has the mass of the strange. Because
this ratio exhibits no dependence on heavy-quark mass, it is just
the Isgur-Wise function $\xi_s(\w)$. The solid curve depicts the
result of fitting the parametrization $s\xi_{NR}$ to the data.
\label{xis_hp}}
\end{minipage}
\end{center}
\end{figure}

The absence of heavy-quark mass dependence of the ratio $h^+/(1+\b^+)$
also means that to a good approximation we have obtained the 
infinite mass result. 
\footnote{The only way $h^+/(1+\b^+)$ could not be the infinite
mass results given it's apparent heavy-quark mass independence is if
option two of the preceding paragraph were realized.} 
Thus, we can combine all of our data for $h^+/(1+\b^+)$ for fixed
light-quark mass but all possible initial and final heavy-quark masses
and call the resulting function an Isgur-Wise function (see
\eq{hpoverrad}). In \fig{xis_hp}, we plot this combined data for the case 
where the mass of the light antiquark is interpolated to the
strange-quark mass.
The resulting Isgur-Wise function, which we denote by $\xi_s(\w)$, is
relevant to the decays $\bstodsl$ and $\bstodssl$ and elastic $B_s$
and $D_s$ scattering off a photon.  The different symbols on the plot
correspond to different initial and/or final heavy-quark mass. The
fact that these different symbols often lie on top of each other and
always appear to lie on the same curve is again confirmation that the
mass dependence of the ratio $h^+/(1+\b^+)$ is very small. The solid
line in \fig{xis_hp} corresponds to a fit of the data to the
parametrization $s\xi_{NR}(\w)$, with $\xi_{NR}$ given in \eq{xinr}.
From this fit we find that this Isgur-Wise function has a slope of
$-\rho^2_s$ at $\w=1$ with
\beq
\rho^2_s=1.2\er{2}{2}\stat\er{2}{1}\syst
\label{rho2s}
\eeq
where the sytematic error was obtained in a way described in 
Ref.\cite{iwprd}.

\begin{figure}[t]
\begin{center}
\leavevmode
\centerline{
\ewxy{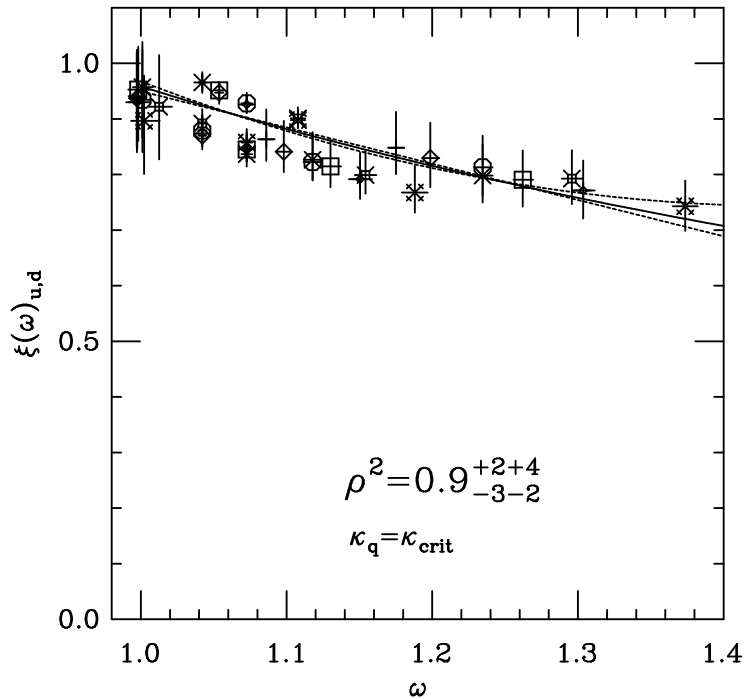}{80mm}
}
\begin{minipage}[t]{\captionwidth}
\caption{The ratio $h^+/(1+\b^+)$ is plotted as a function of $\w$. 
The different symbols correspond to different initial and/or final
heavy-quark masses. All symbols, however, correspond to situations
where the light antiquark is massless. Because
this ratio exhibits no dependence on heavy-quark mass, it is just
the Isgur-Wise function $\xi_{u,d}(\w)$. The solid curve depicts the
result of fitting the parametrization $s\xi_{NR}$ to the data.
\label{xiud_hp}}
\end{minipage}
\end{center}
\end{figure}

In \fig{xiud_hp} we repeat the same exercise with the light-quark mass
extrapolated to the chiral limit. 
The Isgur-Wise function obtained
thus is the one relevant to the decays $\btodl$ and $\btodsl$ and
elastic $B$ and $D$ scattering. This function is traditionally called
``the Isgur-Wise function''. Performing the same fit as in
\fig{xis_hp}, we find that the corresponding slope parameter is
\beq
\rho^2_{u,d}=0.9\er{2}{3}\stat\er{4}{2}\syst
\ .
\label{rho2ud}
\eeq

This slope parameter, as well as the one of \eq{rho2s} is compared
with the predictions of other authors in \tab{rho2comp}.  Our
predictions lie safely above the lower bound of Bjorken\cite{bj} and
below the upper bound of de Rafael and Taron\cite{deraf}. Our results
for $\rho^2_s$ also agree with the lattice result of\cite{bersheso}
although the details and systematics of the two calculations are
different.  The authors of \cite{bersheso} do not quote a value of
$\rho^2_{u,d}$ for vanishing light-quark mass because of their poor
statistics in the chiral limit.
\begin{table}
\centering
\begin{tabular} {|c|c|c|}\hline
Reference & $-\xi'_{u,d}(1)$ & $-\xi'_s(1)$\\
\hline
UKQCD & $0.9\er{2}{3}\stat\er{4}{2}\syst$ & $1.2\er{2}{2}\stat\er{2}{1}\syst$\\
\hline
Bernard, Shen and Soni\cite{bersheso} & & 1.24(26)\stat(26)\syst\\
\hline
de Rafael and Taron\cite{deraf} & $\rho^2<1.42$ &\\
\hline
Close and Wambach\cite{wambach} & 1.40 & 1.64\\
\hline
Neubert\cite{bible} & 0.66(5) &\\
\hline
Voloshin\cite{volosh} & 1.4(3) &\\
\hline
Bjorken\cite{bj} & \multicolumn{2}{c|}{$\rho^2>0.25$}\\
\hline
Blok and Shifman\cite{bs} & $0.35<\rho^2<1.15$ &\\
\hline
H\o gaasen and Sadzikowski\cite{hogasad} & 0.98 & 1.135\\
\hline
Rosner\cite{ros} & 1.59(43) &\\
\hline
Burdman\cite{burd} & 1.08(10) &\\
\hline
Dai, Huang and Jin\cite{dai} & 1.05(20) &\\ \hline
Experiment (see text) & 0.87(12)(20) &\\ \hline
\end{tabular}
\begin{center}
\begin{minipage}[t]{\captionwidth}
\caption{Comparison of our lattice results for $-\xi'_{u,d}(1)$ and 
$-\xi'_s(1)$ to the theoretical predictions of various authors and to
experiment.
\label{rho2comp}}
\end{minipage}
\end{center}
\end{table}

Also for comparison, we quote an average experimental value for the
slope of the Isgur-Wise function compiled by Neubert\cite{neubup} from
very recent results of the ALEPH\cite{aleph} and CLEO\cite{cleo}
Collaborations as well as older data from the ARGUS
Collaboration\cite{argus}:
\beq
\rho^2_{u,d(expt.)}=0.87(12)(20)
\ ,
\eeq
where the second error is theoretical and accounts for the uncertainty
in the size of $1/m_c$ corrections\cite{neubup}.  Agreement with our
result is excellent (see \eq{rho2ud}). Such good agreement, however,
is most certainly coincidental given the size of both the experimental
and lattice errors.

\bigskip

As can be inferred from our determinations of $\rho^2_{u,d}$ and
$\rho^2_s$ (\eqs{rho2ud}{rho2s}) our results suggest that $\rho^2$
decreases slightly as light-quark mass decreases.  Such a decrease is
consistent with one's intuition about the inertia of the light degrees
of freedom.  A very similar trend is also found by H.~H\o gaasen and
M.~Sadzikowski\cite{hogasad}.  In fact, our predictions for $\rho^2$
itself are in excellent agreement with theirs. Their prediction is
based on an improved bag model calculation and is an extension of
earlier work by M.~Sadzikowski and K.~Zalewski\cite{sadzal}. A similar
decrease in slope with spectator-quark mass is observed by F.~E.~Close
and A.~Wambach\cite{wambach} though the values they quote for
$\rho^2_{u,d}$ and $\rho^2_s$ are slightly larger than the ones we
find.

I wish to mention, finally, that J.~Mandula and M.~Ogilvie are in the
process of calculating the Isgur-Wise function using the lattice
version of HQET whose action was given in \eq{shqet}\cite{mandulog}.


\subsection{The Form Factor $h_{A_1}(\w)$}
\label{ha1anal}

We now turn to the subject of $\btodsl$ decays. Here the dominant
form factor is $h_{A_1}$. This form factor is
obtained from the 3-point function
\beq
\sum_{\s x,\s y}\,e^{-i\s q\cdot\s x}e^{-i\s p'\cdot\s y}\,\la
\bar q^R\g_5c^R_s(t_f,\s y)\;\bar c^R\gm\g_5 b^R(t,\s x)\;
\bar b^R_s\g_5 q^R(0,\s 0)\ra
\ ,
\eeq
as $h^+$ was from the 3-point function of \eq{thpthp}.  Here again, we
must study the behavior of the form factor with heavy-quark mass to
obtain the form factor relevant for physical $\btodsl$ decays since
our simulation is performed with heavy-quark masses in the range of
the charm-quark mass. We do so by considering the quantity
\beq
\frac{h_{A_1}(\w)}{1+\b_{A_1}(\w)}\simeq \l(1+\g_{A_1}(\w)\r)\,\xi(\w)
\ ,
\label{ha1overrad}
\eeq
where the equality holds only to leading order in power and radiative
corrections. 

\begin{figure}[t]
\begin{center}
\leavevmode
\centerline{
\ewxy{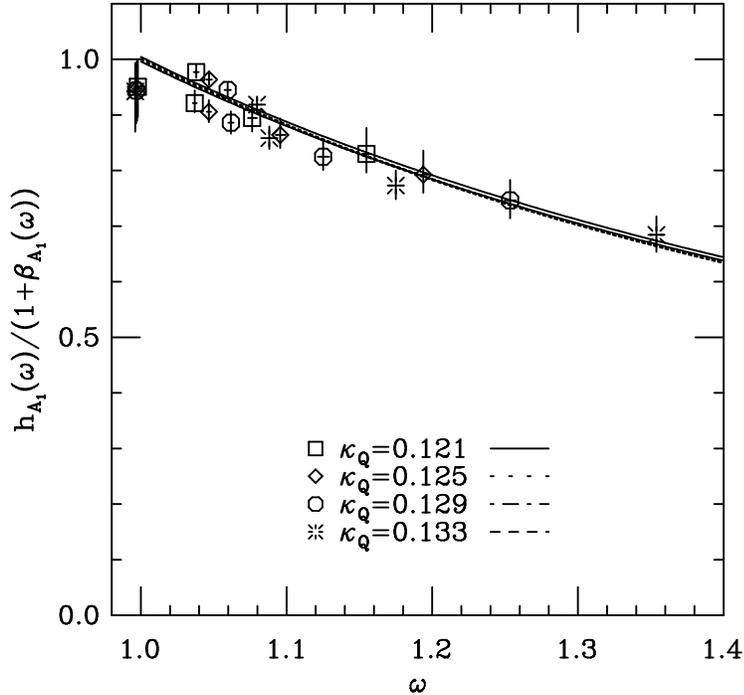}{80mm}
}
\begin{minipage}[t]{\captionwidth}
\caption{The ratio $h_{A_1}/(1+\b_{A_1})$ is plotted as a function
of $\w$ for four different degenerate transitions corresponding to
four values of the heavy-quark mass. The data corresponding to each
transition has it's own symbol as detailed on the plot.  The
different curves result from individually fitting each set of data to
the parametrization $s\xi_{NR}$ (see text). The hopping parameter of
the light, spectator antiquark is $\k_q=0.14144$.
\label{ha1_degen}}
\end{minipage}
\end{center}
\end{figure}

The analysis we perform of the quantity $h_{A_1}/(1+\b_{A_1})$ parallels
the analysis of $h^+/(1+\b^+)$ carried out in \sec{hpanal}. In \fig{ha1_degen}
we plot the ratio $h_{A_1}/(1+\b_{A_1})$ for four degenerate transitions.
Here too we fit the data for each individual degenerate transition to
the parametrization $s\,\xi_{NR}(\w)$, with $\xi_{NR}(\w)$ given in
\eq{xinr}. We find $s=1.00(2)$ and $\rho^2=1.3$ with a statistical
error on the order of $0.3$ for all four data sets. This confirms the
naked eye impression that all four data sets lie on the same curve and
thus indicates that the dependence of the ratio $h_{A_1}/(1+\b_{A_1})$
on heavy-quark mass is very small in the range of masses we are
considering.  Moreover, the fact that the curves on which all of these
points lie are almost the same as those found for $h^+/(1+\b^+)$ 
is a first indication that
the spin component of the heavy-quark symmetry is unbroken in this
case even in the region of the charm-quark mass.

As we did for $h^+/(1+\b^+)$, we can try to quantify the heavy-quark
mass dependence of the ratio $h_{A_1}/(1+\b_{A_1})$ by measuring the
power corrections to this ratio.  If radiative corrections are
neglected, the power corrections to $h_{A_1}$ can be parametrized in
terms of the three univsersal functions $g(\w)$, $\eta(\w)$ and
$g^*(\w)$:
\beq
\g_{A_1}=\frac{\bar\L}{2m_b}\,\l(g(\w)+\frac{\w-1}{\w+1}\l(1-2\,\eta(\w)\r)
\r)+\frac{\bar\L}{2m_c}\,\l(g^*(\w)+\frac{\w-1}{\w+1}\r)
+\ord{\l(\frac{\bar\L}{2m_{c,b}}\r)^2}
\ ,
\label{gsandetadef}
\eeq
where $g(\w)$ is the same form factor that appears in the power
corrections to $h^+$ in \eq{gdef}. Here again, Luke's theorem requires
that the $\ord{\bar\L/2m_{b,c}}$ power corrections to $h_{A_1}$ to
vanish at $\w=1$.  This means that $g^*(\w)$ must vanish at $\w=1$:
\beq
g^*(1)=0
\ .
\label{gs1}
\eeq
$\eta$, however, is unconstrained because it does not appear in the
expression for power corrections at $\w=1$.

\begin{figure}[t]
\begin{center}
\leavevmode
\centerline{
\ewxy{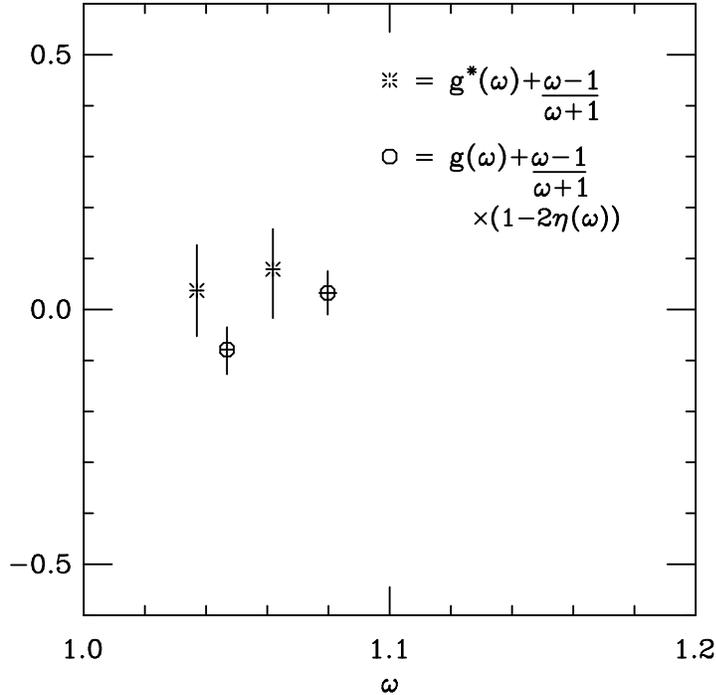}{80mm}
}
\begin{minipage}[t]{\captionwidth}
\caption{Power corrections to the form factor $h_{A_1}$. Plotted are
the functions which multiply the expansion parameters
$\bar\L/(2m_{b,c})$ in \protect{\eq{gsandetadef}}. The hopping parameter of the
light, spectator antiquark is $\k_q=0.14144$.
\label{g_a1_vs_w}}
\end{minipage}
\end{center}
\end{figure}

We extract the functions which multiply the expansion parameters
$\bar\L/(2m_{b,c})$ in \eq{gsandetadef} by taking ratios of the
quantity $h_{A_1}/(1+\b_{A_1})$ at fixed $\w$ and light-quark mass but
different values of the initial or final heavy-quark mass.  We plot
our results in \fig{g_a1_vs_w}.
Both these functions are consistent with 0 and are less than one or
two times $10^{-1}$ in the range of recoils that we can explore. Since
these functions are themselves
multiplied by the small expansion parameters $\bar\L/(2m_c)$ and
$\bar\L/(2m_b)$ which are on the order of 0.2 for the quarks we
simulate, the power corrections to $h_{A_1}$ are very small (on the
order of 4\% or less) in the explored range of recoils. Together
with the information provided by the fits in \fig{ha1_degen}, this
results indicates that power corrections to $h_{A_1}$ are very small
for all $\w$, up to the caveats mentioned in \sec{hpanal} in the
discussion on power corrections to $h^+$.
\footnote{Here too, the manner in which we normalize our data
means that we measure $\g_{A_1}(\w)-\g_{A_1}(1)$ and not
$\g_{A_1}(\w)$. The consequence of this are the same as those
mentioned regarding power corrections to $h^+$.}

\begin{figure}[t]
\begin{center}
\leavevmode
\centerline{
\ewxy{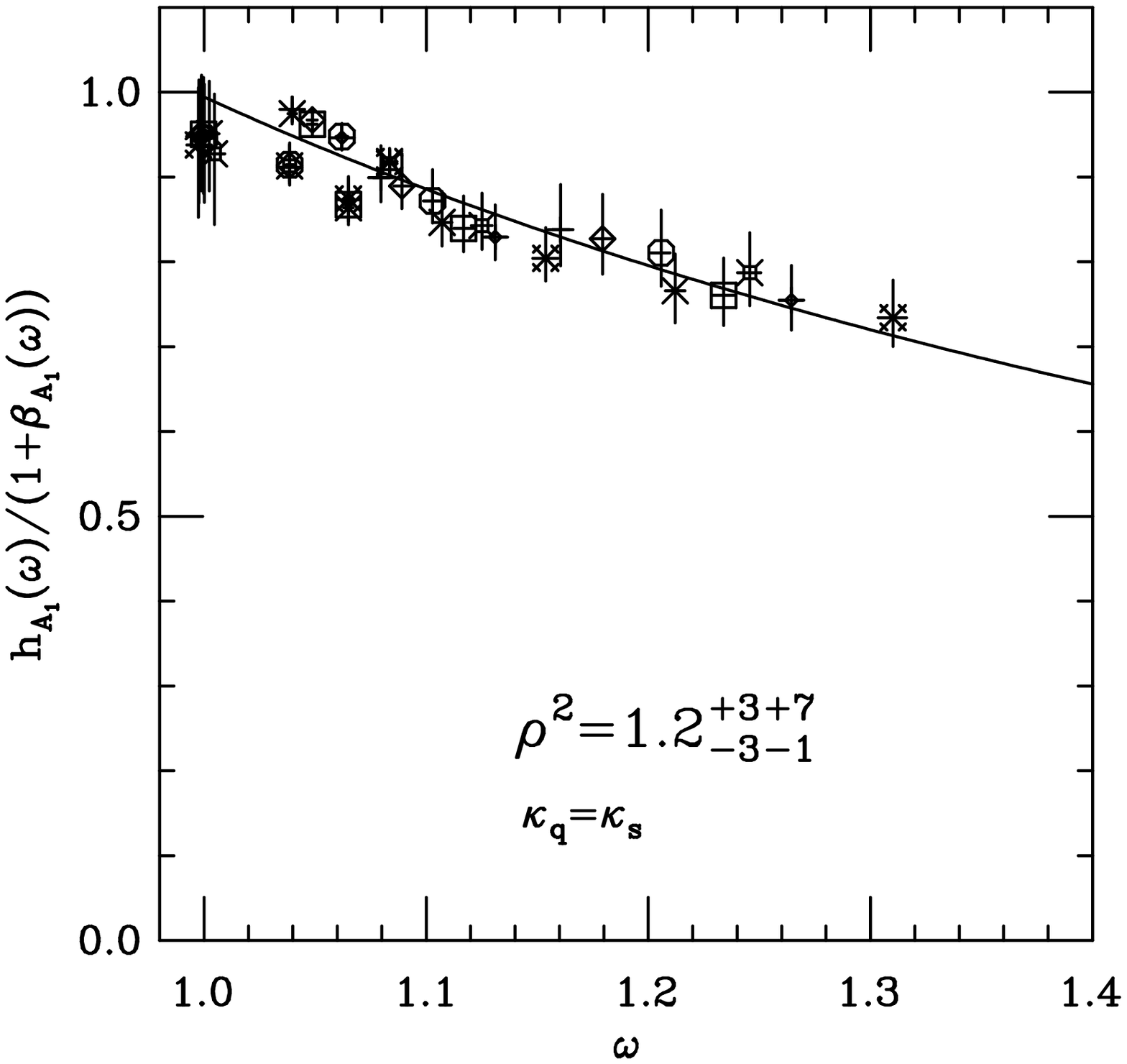}{80mm}
}
\begin{minipage}[t]{\captionwidth}
\caption{The ratio $h_{A_1}/(1+\b_{A_1})$ is plotted as a function of $\w$. 
The different symbols correspond to different initial and/or final
heavy-quark masses. All symbols, however, correspond to situations
where the light antiquark has the mass of the strange. Because
this ratio exhibits no dependence on heavy-quark mass, it is just
the Isgur-Wise function $\xi_s(\w)$. The solid curve depicts the
result of fitting the parametrization $s\xi_{NR}$ to the data.
\label{xis_a1}}
\end{minipage}
\end{center}
\end{figure}

\begin{figure}[t]
\begin{center}
\leavevmode
\centerline{
\ewxy{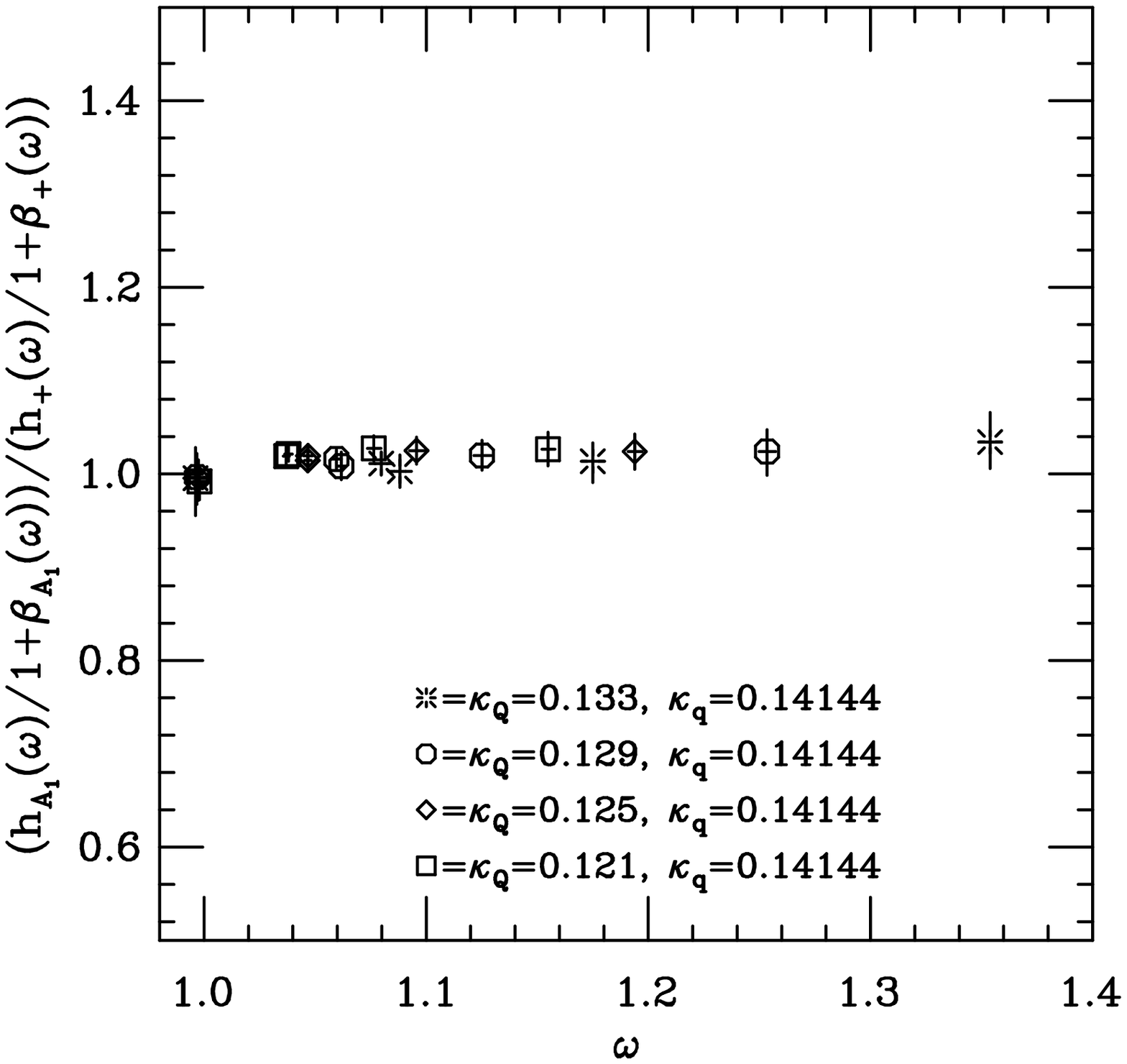}{80mm}
}
\begin{minipage}[t]{\captionwidth}
\caption{The ratio of the Isgur-Wise functions obtained from $h^+$ and
$h_{A_1}$ is plotted as a function of $\w$. Only degenerate
transitions are considered. 
The different symbols correspond to different values of the heavy
quark's mass. All symbols, however, correspond to situations
where the light antiquark has a hopping parameter $\k_q=0.14144$. The
points are very close to one, as they should be if heavy-quark symmetry
is respected. 
\label{ha1hp}}
\end{minipage}
\end{center}
\end{figure}

The fact that $h_{A_1}/(1+\b_{A_1})$ depends very little on
heavy-quark mass means, as it did for the ratio $h^+/(1+\b^+)$, that
our results are to a very good approximation
infinite mass results. So again we can combine all of our data for a given
light-quark mass to obtain the Isgur-Wise functions $\xi_s$ and
$\xi_{u,d}$ which are in principle the same functions as the ones
given by the ratio $h^+/(1+\b^+)$. We plot these form factors as
functions of the four-velocity recoil in
\figs{xis_a1}{xiud_a1}.
\begin{figure}[t]
\begin{center}
\leavevmode
\centerline{
\ewxy{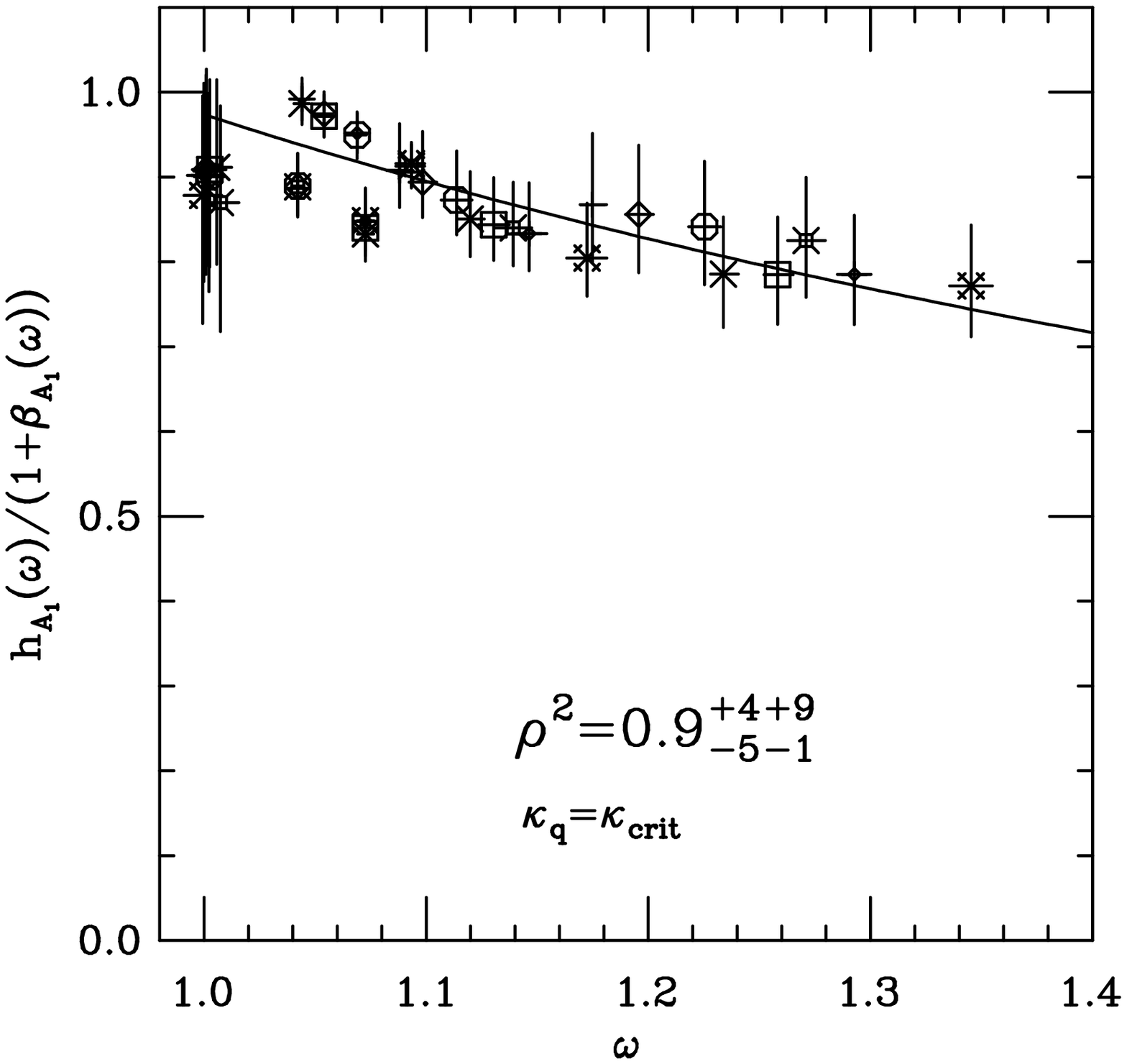}{80mm}
}
\begin{minipage}[t]{\captionwidth}
\caption{The ratio $h_{A_1}/(1+\b_{A_1})$ is plotted as a function of $\w$. 
The different symbols correspond to different initial and/or final
heavy-quark masses. All symbols, however, correspond to situations
where the light antiquark is massless. Because
this ratio exhibits no dependence on heavy-quark mass, it is just
the Isgur-Wise function $\xi_{u,d}(\w)$. The solid curve depicts the
result of fitting the parametrization $s\xi_{NR}$ to the data.
\label{xiud_a1}}
\end{minipage}
\end{center}
\end{figure}
The solid curve in both these plots result from fitting our data to
the parametrization $s\xi_{NR}$. These fits, give the following slope
parameters:
\beq
\rho^2_s=1.2\er{3}{3}\stat\er{7}{1}\syst
\label{rho2s_a1}
\eeq
and
\beq
\rho^2_{u,d}=0.9\er{4}{5}\stat\er{9}{1}\syst
\ .
\label{rho2ud_a1}
\eeq
These slopes are precisely the ones we found for the ratio $h^+/(1+\b^+)$
(see \eqs{rho2s}{rho2ud}). The Isgur-Wise functions obtained
from $h_{A_1}$ and $h^+$ are in fact equal over the whole range
of recoils to within less than 4\%, as can be seen in
\fig{ha1hp} where we plot the ratio $h_{A_1}(1+\b^+)/h^+(1+\b_{A_1})$.

What has emerged is a very consistent picture of $\btodl$ and
$\btodsl$ decays in which heavy-quark symmetry is surprisingly well
satisfied even though the heavy quarks with which we work have masses
in the range of the charm-quark mass. This is in stark contrast with
the results for the decay constant $f_P$ presented in \sec{ldopm}
where we found that corrections to the heavy-quark limit were on the
order of 30\% for these same heavy quarks.  What seems to be happening
here is that the protection from $\ord{\bar\L/2m_{b,c}}$-corrections
that Luke's theorem provides at zero recoil appears to extend over the
full range of recoils so that corrections which one would naively
expect to be on the order of $(\bar\L/2)(1/m_c+1/m_b)\simeq 30-40\%$
for the quarks we are studying turn out to be on the order of a few
percent.


\subsection{Extraction of $V_{cb}$}

\begin{figure}
\begin{center}
\leavevmode
\centerline{
\ewxy{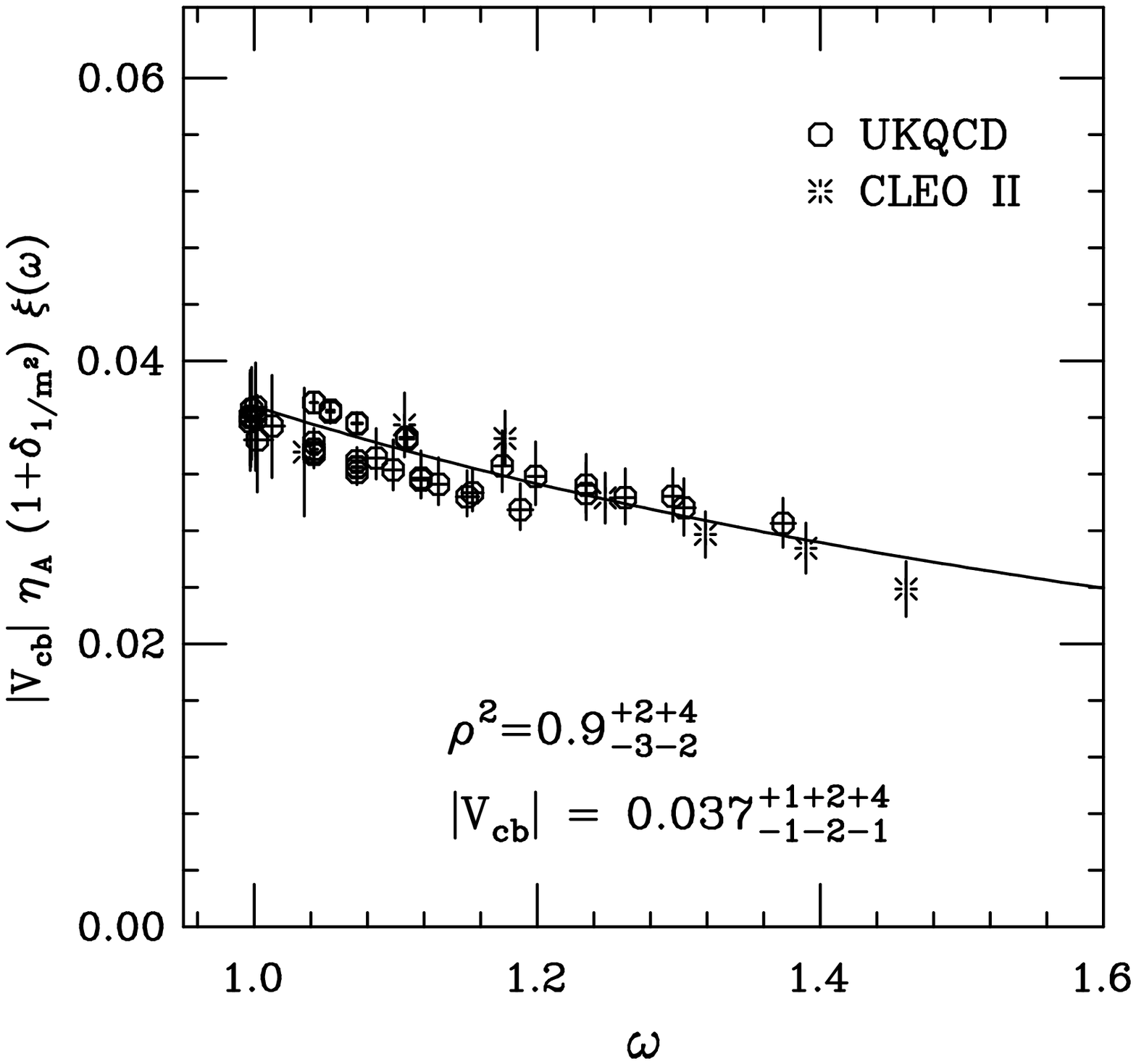}{80mm}
}
\begin{minipage}[t]{\captionwidth}
\caption{$\chi^2$-fit of the expression 
$|V_{cb}|\,\eta_A\,\xi_{u,d}(\w)(1+\d_{1/m^2})$ to the experimental
data for this quantity obtained by the CLEO Collaboration
\protect{\cite{cleo}}. In this
fit, the function $\xi_{u,d}(\w)$ is fixed to our lattice prediction,
i.e.  to $\xi_{NR}$ with $\rho^2$ given by
\protect{\eq{rho2ud}}. The fit parameter is
$|V_{cb}|\,\eta_A\,(1+\d_{1/m^2})$. The UKQCD data and curve are just
those of
\protect{\fig{xiud_hp}} appropriately rescaled by the fit parameter.
The experimental data assumes $\tau_{B^0}=1.53(9) \mbox{ps}$ and
$\tau_{B^+}=1.68(12) \mbox{ps}$.
\label{vcbcleo}}
\end{minipage}
\end{center}
\end{figure}

\begin{figure}
\begin{center}
\leavevmode
\centerline{
\ewxy{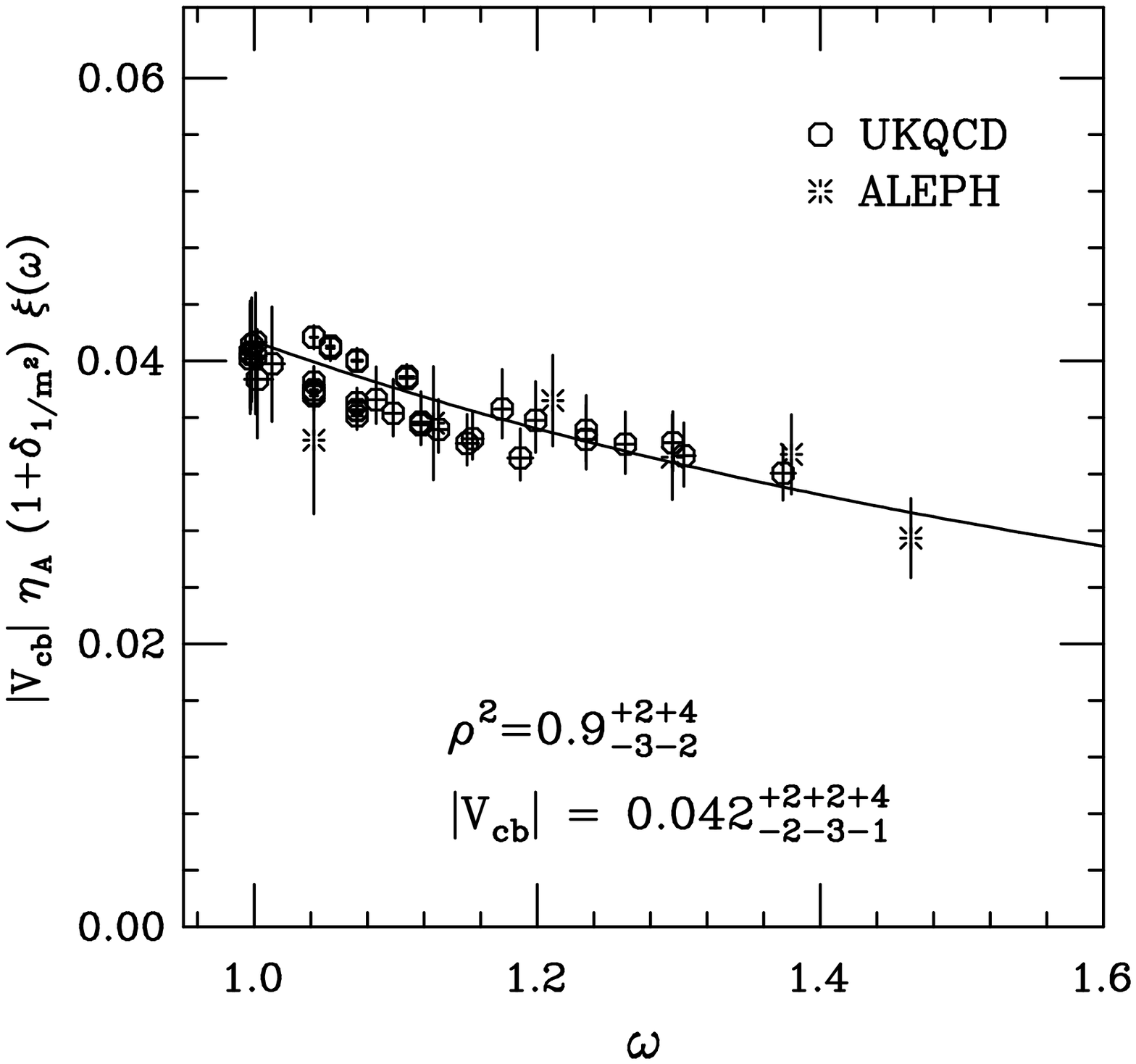}{80mm}
}
\begin{minipage}[t]{\captionwidth}
\caption{Same fit as in \protect{\fig{vcbcleo}} but to
data of ALEPH\protect{\cite{aleph}} which assumes $\tau_{B^0}=1.53(9)
\mbox{ps}$.
\label{vcbaleph}}
\end{minipage}
\end{center}
\end{figure}

\begin{figure}
\begin{center}
\leavevmode
\centerline{
\ewxy{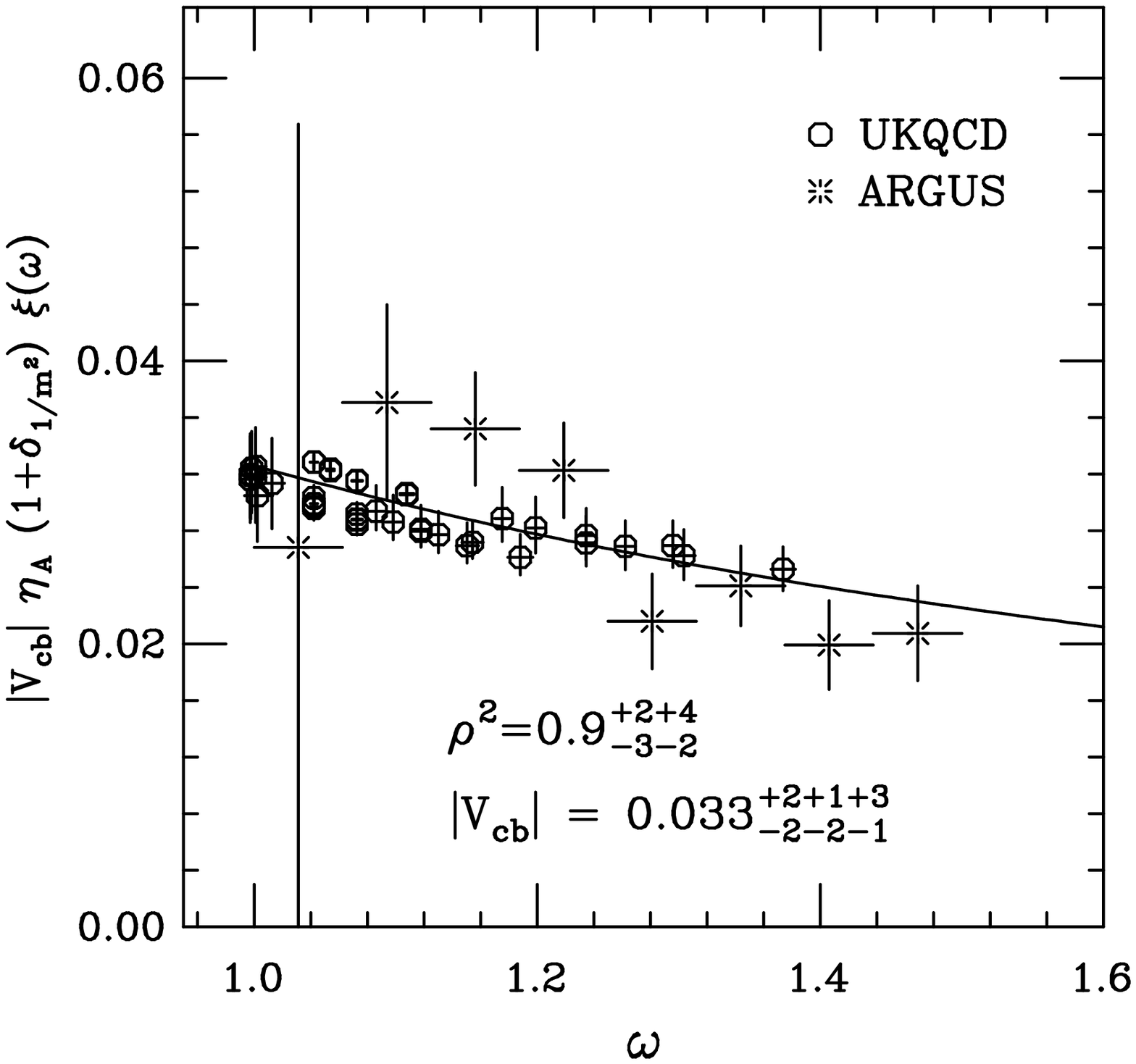}{80mm}
}
\begin{minipage}[t]{\captionwidth}
\caption{Same fit as in \protect{\fig{vcbcleo}} but to
data of the ARGUS Collaboration\protect{\cite{argus}} which assumes
$\tau_{B^0}=1.53(9) \mbox{ps}$.
\label{vcbargus}}
\end{minipage}
\end{center}
\end{figure}

When power corrections and radiative corrections for $\w>1$ are neglected,
the differential decay rate for $\btodsl$ decays is given by
\beq
\frac{1}{\sqrt{\w^2-1}}\frac{d\Gamma(\btodsl)}{d\w}\simeq
\frac{G_F^2|V_{cb}|^2}{4\pi^3}(m_B-m_{D^*})^2m_{D^*}^3\eta_A^2\,(1+\d_{1/m^2})
\,\xi_{u,d}(\w)
\ ,
\label{bdsrate}
\eeq
where $\d_{1/m^2}$ stands for the power corrections to $h_{A_1}$ at
$\w=1$ which have been the object of much controversy
lately\cite{neubup,shifvcb}.  Having determined $\xi_{u,d}$ with
our lattice calculation, the only unknown left in \eq{bdsrate} is the
CKM matrix element $|V_{cb}|$.  Thus, a fit of the theoretical
expression of
\eq{bdsrate} to an experimental measurement of the rate immdediately
yields a measurement of $|V_{cb}|$. This is what we do in
\fig{vcbcleo} where we use very recent data obtained by the CLEO
collaboration\cite{cleo}.
The value of $|V_{cb}|$ that this fit gives is
\beq
|V_{cb}|=0.037\er{1}{1}\er{2}{2}\er{4}{1}\;
\l(\frac{0.99}{\eta_A}\r)\frac{1}{1+\d_{1/m^2}}
\ ,
\eeq
where the first set of errors is due to the experimental uncertainties,
the second set of errors to the statistical errors in our determination
of the Isgur-Wise function and the third to our systematic errors.
A similar fit to ALEPH data\cite{aleph} gives (see \fig{vcbaleph})
%
\beq
|V_{cb}|=0.042\er{2}{2}\er{2}{3}\er{4}{1}\;\sqrt{\frac{1.53{\rm ps}}
{\tau_B}}
\l(\frac{0.99}{\eta_A}\r)\frac{1}{1+\d_{1/m^2}}
\ ,
\eeq
and to ARGUS data\cite{argus} (see \fig{vcbargus}),
%
\beq
|V_{cb}|=0.033\er{2}{2}\er{1}{2}\er{3}{1}\;\sqrt{\frac{1.53{\rm ps}}
{\tau_B}}
\l(\frac{0.99}{\eta_A}\r)\frac{1}{1+\d_{1/m^2}}
\ .
\eeq

These results for $|V_{cb}|$ must not be taken too literally because
the experimental measurements are binned according to slightly biased
estimators of the recoil and because we have neglected small
$1/m_{b,c}$ and radiative corrections for $\w>1$. Neverthless, it is
clear from
\fig{vcbcleo}, \fig{vcbaleph} and \fig{vcbargus} that our 
prediction is consistent with experiment and favors slightly the data of 
CLEO and ALEPH over that of ARGUS.



\subsection{The Form Factor $h_V(\w)$}
\label{hvanal}

We now briefly turn to the form factor $h_V$ defined in
\eq{bds}. This form factor is interesting because it is not
protected by Luke's theorem as are $h^+$ and $h_{A_1}$. This
means that one would expect power corrections to this form factor
to be more in line with naive expectations.

We have very prelimary results for the form factor $h_V$ and it does
indeed seem to exhibit power corrections on the order of 20 to 40\%
depending on the heavy-quarks considered. It is unclear, however, what
fraction of these corrections are true $1/m_{b,c}$-corrections and
what fraction are $am_Q$ discretization errors. The problem here is
that we cannot subtract these discretization errors as we did for
$h^+$ and $h_{A_1}$ (see Ref. \cite{iwprd} for details), because there
is no normalization condition for $h_V$.\footnote{For the case of $h^+$
and $h_{A_1}$, Luke's theorem guaranteed the these form factors would
be 1 at $\w=1$ up to calculable perturbative corrections and small
non-perturbative corrections $\ord{\l(\bar\L/m_{b,c}\r)^2}$.}

So instead of presenting results which may suffer from large discretization
errors, we will present a framework in which the form factor $h_V$ may
be analyzed once these errors are controlled. 
As we did for these two other form factors, we should 
study $h_V$ with the radiative corrections taken out:
\beq
\frac{h_V(\w)}{1+\b_V(\w)}\simeq (1+\g_V(\w))\,\xi(\w)
\ ,
\eeq
where $\g_V$ labels power corrections. These corrections can be
parametrized by the three universal functions $g$, $g^*$ and $\eta$
that we encountered in \sec{hpanal} and \sec{ha1anal}. In the absence of
radiative corrections, 
\beq
\g_V=\frac{\bar\L}{2m_b}\,\l(g(\w)+(1-2\,\eta(\w))\r)
+\frac{\bar\L}{2m_c}\,\l(g^*(\w)+1\r)
+\ord{\l(\frac{\bar\L}{2m_{c,b}}\r)^2}
\ .
\label{gammav}
\eeq
Then, by considering ratios of the quantity $h_V/(1+\b_V)$ for fixed
$\w$ and light-quark mass, but different heavy-quark masses, we would
extract the functions $g(\w)+(1-2\,\eta(\w))$ and $g^*(\w)+1$. We
would expect both these functions to be of $\ord{1}$ since there is no
symmetry, here, which forbids the appearance of power corrections. It
is interesting to note that this expectation is consistent with the
statement that $g^*(\w)$ must vanish at $\w=1$ (see \eq{gs1}).

Having obtained these functions, we would combine them with the
results for the power corrections to $h^+$ and $h_{A_1}$ of
\figs{g_p_vs_w}{g_a1_vs_w} and solve the resulting system of equations
at fixed $\w$ for the subleading universal
functions $g$, $g^*$ and $\eta$.

We have actually carried out this whole procedure on our preliminary
data and find results for $g$, $g^*$ and $\eta$ which are very much in
line with what one can infer from sumrule calculations\cite{bible}.



\section{Penguins on the Lattice
\protect{\footnote{Much the UKQCD data presented in this section
have appeared in Ref.\protect{\cite{brian}} or will appear in its revised
version.}}}
\label{bsgsec}

\begin{figure}[t]
\begin{center}
\leavevmode
\centerline{
\setlength{\epsfxsize}{450pt}\epsfbox[0 450 640 675]{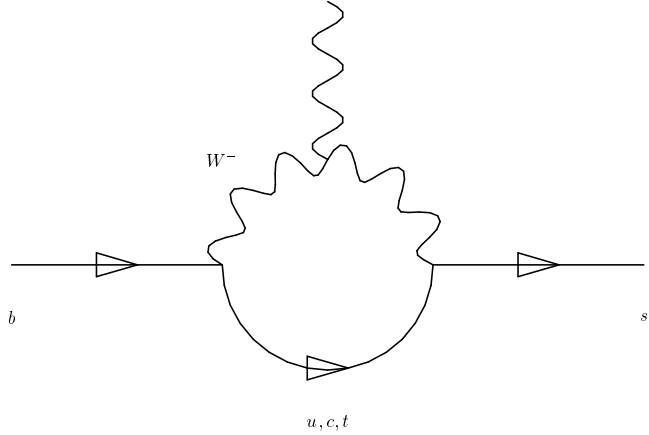}
}
\begin{minipage}[t]{\captionwidth}
\caption{Example of a penguin diagram that contributes to the 
decay $b\to s\g$.
\label{penguin}}
\end{minipage}
\end{center}
\end{figure}

We now turn to the study of $\bksg$ decays. These decays occur through
the quark level process $b\to s\g$ depicted in \fig{penguin} which 
mediated by a flavor-changing-neutral-current (FCNC).
Such FCNC processes are interesting because they are fobidden at tree level
in the Standard Model and only occur at one or higher-loop order.
Their study therefore provides a means for testing the details
of the Standard Model. And because even at lowest order they are sensitive
to the presence of new particles that may appear in the loops
of their diagrams, they may even give us a handle on physics beyond
the Standard Model at comparatively low energies. In fact, bounds on
the $b\to s\g$ branching ratio have already been used to place
constraints on supersymmetric as well as non-supersymmetric extensions
to the Standard Model. A comprehensive review of these results can be
found in Ref.\cite{hewett}.

Another interesting aspect of $b\to s\g$ decays is that they may permit
a measurement of the poorly determined CKM matrix element $V_{ts}$.

Experimentally, $b\to s\g$ decays are difficult for the same reasons
that make them interesting theoretically. They occur at one loop and
their rate is suppressed by two powers of the Fermi constant $G_F$:
they are rare decays. Nevertheless, the CLEO Collaboration has been able to
measure the branching ratio for the exclusive channel $\bksg$.  It
found that this decay has a branching ratio\cite{cleobksg}
\beq
{\cal BR}\l(\bksg\r)=(4.5\pm 1.5\pm 0.9)\times 10^{-5}
\ .
\label{cleobksg}
\eeq

In order to compare this very pretty experimental result with the
predictions of the Standard Model, one has to compute the
long-distance contributions of the strong interaction which are, of
course, non-perturbative. These long-distance contributions are given
by the hadronic matrix element $\la K^*|\,\bar
s\sigma_{\mu\nu}q^\mu(1+\g_5)b\,|B\ra$ which can be parametrized by
three form factors\cite{berhsison}
\beq
\la K^*(k,\epsilon)|\,\bar s\sigma_{\mu\nu}q^\nu\frac{(1+\g_5)}{2}b\,|B(p)\ra
=\sum_{i=1}^3 C^i_\mu T_i(q^2) ,
\end{equation} 
where $k$ and $\epsilon$ are the momentum and polarization vectors
of the $K^*$; $p$ is the momentum of the $B$; $q=p-k$; and
\begin{eqnarray}
C^{1}_\mu & = & 
2 \varepsilon_{\mu\nu\lambda\rho} \epsilon^\nu p^\lambda k^\rho, \\
C^{2}_\mu & = & 
i\epsilon_\mu(m_B^2 - m_{K^*}^2) - i\epsilon\cdot q (p+k)_\mu, \\
C^{3}_\mu & = & 
i\epsilon\cdot q 
\left( q_\mu - \frac{q^2}{m_B^2-m_{K^*}^2} (p+k)_\mu \right).
\end{eqnarray}
On the lattice, these form factors are obtained from a 3-point
function in very much the same way as were the semi-leptonic form
factors of \sec{semi-lep}. The viability of this particular
calculation was first demonstrated by Bernard, Hsieh and Soni in
Ref.\cite{berhsison}.

As the photon emitted is on-shell, the form factors need only be
evaluated at $q^2{=}0$.  In this limit,
\begin{equation}
T_2(q^2{=}0)  =  T_1(q^2{=}0) ,
\label{t1eqt2}
\end{equation} 
and the coefficient of $T_3(q^2{=}0)$ vanishes. Hence, the branching
ratio can be expressed in terms of a single number which we take
to be $T_1(q^2{=}0)$.

As in our studies of leptonic and semi-leptonic decays of $B$ mesons,
the simulation is performed with the four heavy quarks listed in
\tab{heavy}. Since these quarks have masses around that of the charm,
results have to be extrapolated to $m_b$.  Here, however, the
dependence of the relevant form factors on heavy-quark mass is not as
straighforward as it was for leptonic decay constants and
semi-leptonic form factors because the light degrees of freedom in
this decay have momenta comparable to the mass of the $b$-quark in
large sections of phase space.  Nevertheless, in a region around
$q^2_{max}=(m_B-m_{K^*})^2$ heavy-quark symmetry will apply. In that
region, one finds that $T_1(q^2)$ and $T_2(q^2)$ scale according
to\cite{iwbsg}:
\beq
T_1\l(q^2\simeq q^2_{max};m_P;m_{K^*}\r)=
a_0 \sqrt{m_P}\times [ \alpha_s(m_P) ]^{-2/\beta_0}
 \left(1 + \frac{a_1}{m_P} + \frac{a_2}{m_P^2} + \dots\right)
\label{t1q2maxscale}
\eeq
and
\beq
T_2\l(q^2\simeq q^2_{max};m_P;m_{K^*}\r)=
b_0 \frac{1}{\sqrt{m_P}}\times [ \alpha_s(m_P) ]^{-2/\beta_0}
 \left(1 + \frac{b_1}{m_P} + \frac{b_2}{m_P^2} + \dots\right)
\ ,
\label{t2q2maxscale}
\eeq
with the same one-loop running coupling $\a_s$ and the same $\b_0$ as
for the leptonic case (see comments after \eq{phimplim}).  

\begin{figure}[t]
\begin{center}
\leavevmode
\centerline{
\ewxy{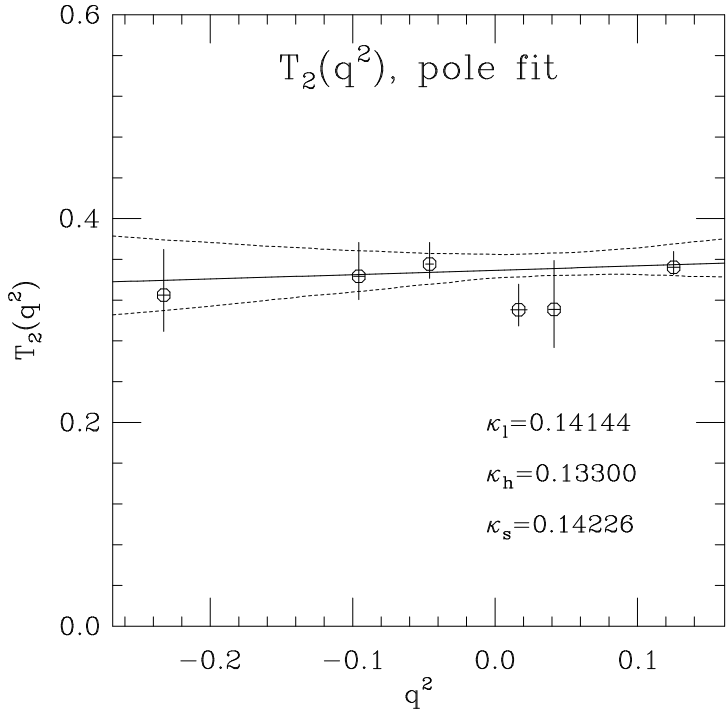}{80mm}
}
\begin{minipage}[t]{\captionwidth}
\caption{$T_2$ plotted as a function of $q^2$. Even though $T_2$ is
consistent with pole dominance (solid line), 
it is clear that $T_2$ could also be constant. The dotted lines represent
the 68\% confidence levels of the fit to the pole form.
\label{t2const}}
\end{minipage}
\end{center}
\end{figure}

The problem
now is to find a way to use these scaling relations to get
$T_1\l(q^2{=}0;m_B;m_{K^*}\r)$.  The first approach is to:
\footnote{This approach can also, of course, be followed
with $T_1$ instead of $T_2$.}
\begin{enumerate}
\item 
extrapolate the direct measurements of $T_2\l(q^2=
q^2_{max};m_P;m_{K^*}\r)$ to $m_P=m_B$;
\item
use the results for
this form factor at fixed $m_P\simeq m_D$ to make an educated guess
about its behavior in $q^2$ at $m_B$;
\item
use this behavior to extrapolate the value of 
$T_2\l(q^2=q^2_{max};m_B;m_{K^*}\r)$ obtained in step 1 to $q^2=0$.  
\end{enumerate}
This method has the
advantage that it only requires knowledge of $T_2$'s $q^2$ dependence
at $m_B$. Its disadvantage, however, is that one has to extrapolate
$T_2$ over a wide range of momentum transfers from
$q^2=q^2_{max}$ to $q^2=0$. Since one does not exactly know the
functional behavior of $T_2$ in $q^2$, this extrapolation leads to rather 
large uncertainties.  Bernard, Hsieh and Soni make a rather good
case for the use of pole dominance\cite{berhsisonprl}, but
it is difficult to exclude the possibility that $T_2$ might
be a constant on the basis of the data alone (see \fig{t2const}
taken from Ref.\cite{brian}). 

\begin{figure}[t]
\begin{center}
\leavevmode
\centerline{
\ewxy{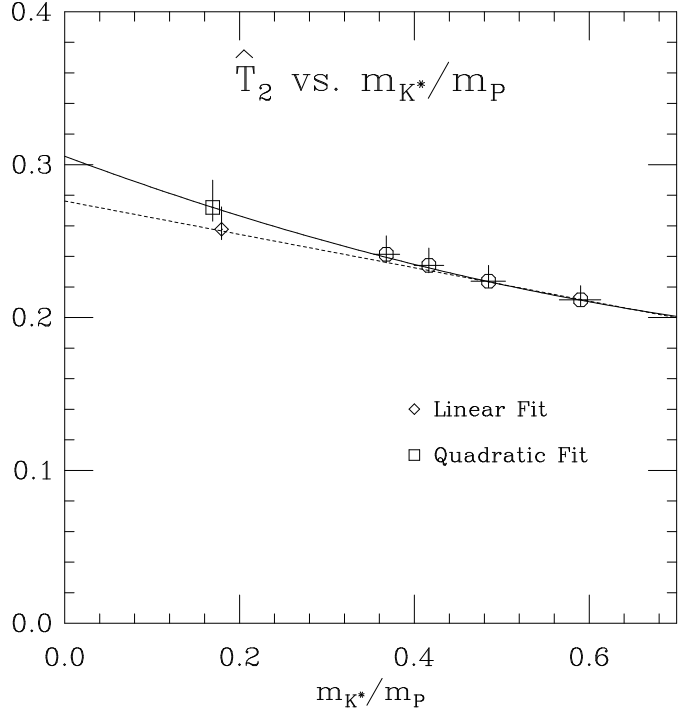}{80mm}
}
\begin{minipage}[t]{\captionwidth}
\caption{$\hat T_2(m_P)$ is extrapolated lineary (dotted line) 
and quadratically (solid curve) to $m_P=m_B$.
\label{t2q2maxextrap}}
\end{minipage}
\end{center}
\end{figure}

Step 1 is performed by extrapolating the quantity
\beq
{\hat T}_2(m_P)=
T_2(q^2_{max}) 
\sqrt{m_P \over m_B} 
\left({\alpha_s(m_P)\over\alpha_s(m_B)}\right)^{2/\beta_0},
\label{t2hat}
\eeq
linearly and quadratically in $1/m_P$ to $1/m_B$ (see 
\eq{t2q2maxscale}). This extrapolation is shown in \fig{t2q2maxextrap}. 
It yields
\begin{equation}
T_2(q^2{=}q^2_{max};m_B;m_{K^*}) = 0.269^{+17}_{-9}\pm{0.011}\ ,
\label{t2q2maxres}
\end{equation}
taking the quadratic fit as the best 
estimate and the difference between the central values of the
linear and quadratic fits as an estimate of the sytematic error
(the second error in \eq{t2q2maxres}). 

Now, if pole behavior is assumed for $T_2(q^2;m_B;m_{K^*})$, this result for
$T_2(q^2{=}q^2_{max};m_B;m_{K^*})$ implies (see \eq{t1eqt2})
\begin{equation}
\label{t20pole}
T_1(q^2{=}0;m_B;m_{K^*})=
T^{\mbox{\scriptsize\it{pole}}}_2(q^2{=}0;m_B;m_{K^*}) 
= 0.112^{+7}_{-7}\mbox{}^{+16}_{-15},
\end{equation}
where the first error is statistical and the second is the systematic
error obtained by combining the variation of the pole mass within its
bounds and the systematic error from \eq{t2q2maxres}. The particle
exchanged here is the $1^+$, $B_{s1}$ state whose mass has not
yet been measured. The authors of Ref.\cite{brian} are nevertheless
able to estimate this mass to be $m_{B_{s1}}=5.74\pm0.21\gev$
by using HQET.

If one assumes, on the other hand, that $T_2(q^2;m_B;m_{K^*})$ is
constant, then $T_1(q^2{=}0;m_B;m_{K^*})$ is immediately given by
\eq{t2q2maxres}.

\bigskip
The second approach to obtaining $T_1(q^2{=}0;m_B;m_{K^*})$
is to translate the scaling relations of
\eqs{t1q2maxscale}{t2q2maxscale} to scaling relations for
$T_{1,2}\l(q^2=0;m_P;m_{K^*}\r)$. This requires an
assumption about the $q^2$ dependence of the form factors for all
$m_P$.  As was pointed out by As.~Abada in Ref.\cite{abada}, it is
inconsistent to assume that both form factors behave according to a
pole dominance form. Indeed, if one combines this assumption with the
scaling relations of \eqs{t1q2maxscale}{t2q2maxscale}, one finds that
the form factors scale at $q^2{=}0$ as $T_1(0)\sim m_P^{-1/2}$ and
$T_2(0)\sim m_P^{-3/2}$ in clear contradiction with the fact that
$T_1(0)=T_2(0)$ (\eq{t1eqt2}). 
Thus, only one of the form factors can
obey pole dominance. If one assumes that $T_1(q^2)$ does, which is consistent
with $T_2(q^2)$ being constant, 
then
\eq{t1q2maxscale} implies that $T_1(0;m_P)$ scales according to:
\beq
T_1\l(q^2{=}0;m_P;m_{K^*}\r)=
c_0 m_P^{-1/2}\times [ \alpha_s(m_P) ]^{-2/\beta_0}
 \left(1 + \frac{c_1}{m_P} + \frac{c_2}{m_P^2} + \dots\right)
\ .
\label{t10scalepole}
\eeq
If, on the other hand, pole dominance for $T_2(q^2)$ is assumed, which
is consistent with $T_1(q^2)$ having a dipole form, then by
\eqs{t2q2maxscale}{t1eqt2} $T_1(0;m_P)$ scales as:
\beq
T_1\l(q^2{=}0;m_P;m_{K^*}\r)=
d_0 m_P^{-3/2}\times [ \alpha_s(m_P) ]^{-2/\beta_0}
\left(1 + \frac{d_1}{m_P} + \frac{d_2}{m_P^2} + \dots\right)
\ .
\label{t10scaledipole}
\eeq
To obtain these scaling relations one has 
to expand the pole masses and $q^2_{max}$ in inverse powers of $m_P$
which is reasonable in the limit of large $m_P$.
The advantage of this second approach is that it does not require
an extrapolation over a wide range of momentum transfers. Its
disadvantage is that it requires one to assume that the form
factors has a certain form for all $m_P$ in the range from
$\sim m_D$ to $m_B$.

\begin{figure}[t]
\begin{center}
\leavevmode
\centerline{
\ewxy{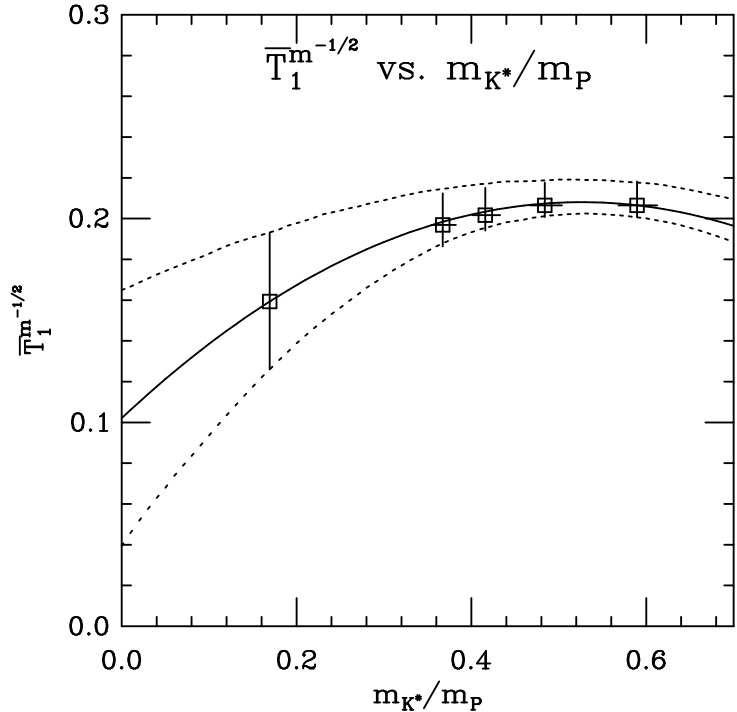}{80mm}
}
\begin{minipage}[t]{\captionwidth}
\caption{The solid curve represents the quadratic extrapolation
of ${\hat T}_1^{m^{-1/2}}(m_P)$ to $m_{K^*}/m_B$. The dotted
curves correspond to the statistical errors on the fit.
\label{t10polefig}}
\end{minipage}
\end{center}
\end{figure}

To implement the scaling relation of \eq{t10scalepole}, we construct
the quantity 
\beq
{\hat T}_1^{m^{-1/2}}(m_P)=
T_1(q^2{=}0) 
\sqrt{m_P \over m_B} 
\left({\alpha_s(m_P)\over\alpha_s(m_B)}\right)^{2/\beta_0},
\label{t1hatoh}
\eeq
as we did for the scaling of $T_2$. Extrapolating ${\hat
T}_1^{m^{-1/2}}(m_P)$ quadratically in $1/m_P$ to $1/m_B$ then gives
(see \fig{t10polefig})
\beq
T_1\l(q^2{=}0;m_B;m_{K^*}\r)=0.159\err{34}{33}
\ .
\label{t10pole}
\eeq
\noindent
There is a problem here, however. One can also obtain the scaling
relation of \eq{t10scalepole} by combining \eqs{t1eqt2}{t2q2maxscale}
with the assumption that $T_2(q^2,m_P)$ is a constant function of
$q^2$ for all $m_P$ in the range of $m_D$ to $m_B$. But if one does
this, one has to conclude that ${\hat T}_2(m_P)= {\hat
T}_1^{m^{-1/2}}(m_P)$ in contradiction with the results of
\figs{t10polefig}{t2q2maxextrap}. So it appears that the preliminary
data of the
UKQCD Collaboration is inconsistent with the scaling relation of
\eq{t10scalepole} and therefore with the assmptions that led to it.

\begin{figure}[t]
\begin{center}
\leavevmode
\centerline{
\ewxy{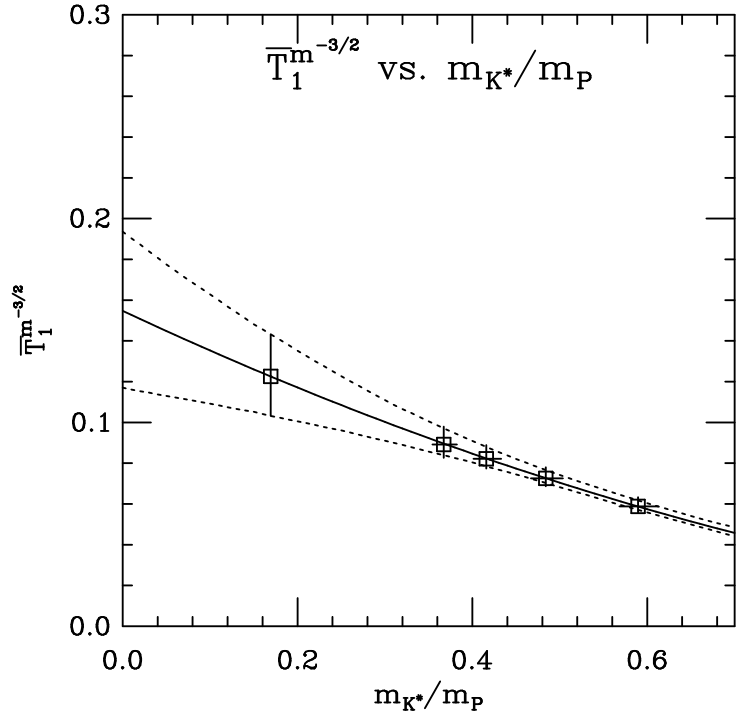}{80mm}
}
\begin{minipage}[t]{\captionwidth}
\caption{The solid curve represents the quadratic extrapolation
of ${\hat T}_1^{m^{-3/2}}(m_P)$ to $m_{K^*}/m_B$. The dotted
curves correspond to the statistical errors on the fit.
\label{t10dipolefig}}
\end{minipage}
\end{center}
\end{figure}

The extrapolation of $T_1\l(q^2{=}0;m_P;m_{K^*}\r)$ according
to \eq{t10scaledipole} is performed, on the other hand, 
by extrapolating the quantity
\beq
{\hat T}_1^{m^{-3/2}}(m_P)=
T_1(q^2{=}0) 
\l({m_P \over m_B}\r)^{3/2}
\left({\alpha_s(m_P)\over\alpha_s(m_B)}\right)^{2/\beta_0}
\label{t1hatth}
\eeq
quadatrically in $1/m_P$ as shown in 
\fig{t10dipolefig}. This extrapolation gives
\beq
T_1\l(q^2{=}0;m_B;m_{K^*}\r)=0.123\err{21}{19}
\ .
\label{t10dipole}
\eeq

In \tab{t10sum}, I summarize the different values for 
$T_1\l(q^2{=}0;m_B;m_{K^*}\r)$ obtained
above. I add to these results a systematic error of
$\a_sam_Q\simeq 8\%$, where $m_Q$ is the bare mass of our heaviest
heavy-quark, to account for possible discretization errors
as discussed in the section on Symanzik improvement.
\begin{table}
\centering
\begin{tabular}{|c|c|}\hline
method & $T_1\l(q^2{=}0;m_B;m_{K^*}\r)$ \\
\hline
$T_2(q^2_{max})\sim m^{-1/2}$, $T_2(q^2;m_B)$ constant 
& $0.269^{+17}_{-9}\pm{0.024}$\\
\hline
$T_2(q^2_{max})\sim m^{-1/2}$, $T_2(q^2;m_B)$ pole 
& $0.112^{+7}_{-7}\pm{0.018}$\\
\hline
$T_1(0)\sim m_P^{-1/2}$ ($T_2(q^2;m_P)$ constant)
& $0.159\err{34}{33}\pm{0.015}$\\
\hline
$T_1(0)\sim m_P^{-3/2}$ ($T_2(q^2;m_P)$ pole)
& $0.123\err{21}{19}\pm{0.010}$\\
\hline
\end{tabular}
\begin{center}
\begin{minipage}[t]{\captionwidth}
\caption{Preliminary UKQCD results for $T_1\l(q^2{=}0;m_B;m_{K^*}\r)$
as obtained from the various assumptions described in the text.
\label{t10sum}}
\end{minipage}
\end{center}
\end{table}
The consistency of the $T_2$-pole results plus the fact that the
scaling relation of \eq{t10scalepole} appears to be inconsistent with
the preliminary lattice results of the UKQCD Collaboration 
together seem to indicate that a value for
$T_1(0,m_B)$ of around 0.11-0.12 is favored. However, since these
results are still preliminary and do not belong to me, I shall not venture
a final number. Instead, 
in \tab{t10comp} I provide the results obtained
by other lattice groups for comparison.  
\begin{table}
\centering
\begin{tabular} {|c|c|c|c|c|}\hline
Ref.& Action & $\beta$ & \multicolumn{2}{c|}{method}\\
\hline
APE\cite{apebksg} & Clover & 6.0 & $T_1(0)\sim m_P^{-1/2}$ 
& $T_1(0)\sim m_P^{-3/2}$\\ 
\cline{4-5}
& & & 0.21(4)(2) & 0.12(2)(3) \\ \hline
BHS\cite{berhsisonprl} & Wilson & 6.3 & 
\multicolumn{2}{c|}{$T_2(q^2_{max})\sim m^{-1/2}$, $T_2(q^2;m_B)$ pole}\\
\cline{4-5}
& & & \multicolumn{2}{c|}{$0.10\pm.01\pm.03$} \\ \hline
LANL\cite{lanl} & Wilson & 6.0 & $T_1(0)\sim m_P^{-1/2}$ 
& $T_1(0)\sim m_P^{-3/2}$\\ 
\cline{4-5}
& & & 0.25(2)& 0.09(1) \\ \hline
\end{tabular}
\begin{center}
\begin{minipage}[t]{\captionwidth}
\caption{Results for $T_1\l(q^2{=}0;m_B;m_{K^*}\r)$ 
obtained by other lattice groups making
use of a variety of methods described in the text and
summarized in \protect{\tab{t10sum}}. All of these results
are preliminary except for those of BHS. The LANL group also gets,
using the method of BHS, $T_1\l(q^2{=}0;m_B;m_{K^*}\r)=0.10(1)$.
\label{t10comp}}
\end{minipage}
\end{center}
\end{table}

Finally, for comparison with experiment, it is useful to convert
$T_1(0)$ into a value for the dimensionless hadronization ratio
\beqa
R_{K^*} &=& \frac{\wid{\bksg}}{\wid{b\to s\g}}\nonumber\\
&=& 4\l(\frac{m_B}{m_b}\r)^3\l(1-\frac{m_{K^*}^2}{m_B^2}\r)^3
\; |T_1(q^2{=}0)|^2
\ .
\label{rks}
\eeqa
Many of the theoretical uncertainties which arise in relating the
amplitudes for these decays to their branching ratios cancel in this
ratio. The values for $R_{K^*}$ obtained from the results for $T_1(0)$
summarized in \tab{t10sum} are given in \tab{rkssum}. These values for
$R_{K^*}$ are independent of whether one considers $B^0\to K^{*0}\g$
or $B^\pm\to K^{*\pm}\g$ decays. I have chosen $m_b=4.87\gev$ for
consistency with Ref.\cite{cleobsg} and used the 1994 Particle Data
Book\cite{pdg94} for all other masses. Also given in \tab{rkssum} is
the experimental result. This number is obtained by taking the ratio of CLEO's
measurement of the branching ratio $\br{\bksg}$ given in
\eq{cleobksg} to their measurement of the inclusive branching ratio
$\br{b\to s\g}=2.32\pm0.51\pm0.29\pm0.32$\cite{cleobsg}. 
\begin{table}
\centering
\begin{tabular}{|c|c|}\hline
method & $R_{K^*}$ \\
\hline
$T_2(q^2_{max})\sim m^{-1/2}$, $T_2(q^2;m_B)$ constant 
& $(34\er{4}{2}\pm3\pm5)\%$\\
\hline
$T_2(q^2_{max})\sim m^{-1/2}$, $T_2(q^2;m_B)$ pole 
& $(5.9\err{0.7}{0.7}\pm{1.9})\%$\\
\hline
$T_1(0)\sim m_P^{-1/2}$ ($T_2(q^2;m_P)$ constant)
& $(12\er{5}{5}\pm{2})\%$\\
\hline
$T_1(0)\sim m_P^{-3/2}$ ($T_2(q^2;m_P)$ pole)
& $(7.1\errr{2.4}{2.2}\pm{1})\%$\\
\hline
CLEO (see text) & $(19\pm 13)\%$
\\ \hline
\end{tabular}
\begin{center}
\begin{minipage}[t]{\captionwidth}
\caption{Comparison with experiment of preliminary UKQCD results for $R_{K^*}$ 
obtained from the results for 
$T_1\l(q^2{=}0;m_B;m_{K^*}\r)$ of \protect{\tab{t10sum}}.
\label{rkssum}}
\end{minipage}
\end{center}
\end{table}
Both experimental and lattice uncertainties are so large at this
early stage that it is difficult to draw any firm conclusion from 
a comparison. 

Before ending this discussion of radiative $B$ decays, I would like to
make a brief comment on corrections to the heavy quark limit.  
\fig{t2q2maxextrap} shows that ${\hat T}_2$ suffers
13\% corrections at the scale of the $B$ and 27\% corrections at
$m_D$.  This is very much in line with what we found for leptonic
decays.


\section{Conclusion}
\label{concl}

Lattice QCD studies are already providing information about the strong
interaction effects in the weak decays of $B$-mesons which is
of fundamental phenomenological and
theoretical importance. 

As far as phenomenology is concerned, we have seen that lattice studies of
leptonic decays of $B$-mesons have lead to predictions for the decay
constant $f_B$ required for describing $B-\bar B$ mixing as well as
non-leptonic decays in factorization approximations\cite{ref108109}.
These predictions are well summarized by the statement
\beq
f_B=180\pm 40\mev
\ .\eeq
We have also seen that lattice simulations can be used to determine
the form factors required for guiding the extraction of the CKM
parameter $|V_{cb}|$ from experimental measurements of the
differential decay rate for $\btodsl$ decays. In Ref. \cite{iwprd}
these form factors are further used to predict various semi-leptonic
$B$-meson decay rates.  Finally, we saw that the lattice is beginning
to make predictions for the form factors relevant to the rare decay
$B\to K^*\g$ and for the corresponding hadronization ratio $R_{K^*}$.
Because rare decays are
sensitive, low-energy probes for physics beyond the standard model,
these predictions are very important. 

\medskip

On the theoretical side, the fact that quark masses are adjustable in
lattice calculations enables one to trace out precisely the
dependence of various quantities on heavy-quark mass and hence probe
the range of applicability of heavy-quark symmetry and test the validity
of HQET. When studying
leptonic decay constants we found that power
corrections to the heavy-quark limit were on the order of 10-15\% at the scale
of the $b$-quark and 30-40\% at $m_c$. This is significantly
larger than one would expect on the grounds that these corrections
are proportional to $\lqcd/m_b\simeq 5\%$ and 
$\lqcd/m_c\simeq 17\%$, respectively, but is consistent with other
theoretical determinations\cite{bib104138139}.

When studying semi-leptonic $\btodl$ and $\btodsl$ decays, on
the other hand,  we found 
that the form factors $h^+$ and $h_{A_1}$ suffer power
corrections which are much smaller than one would expect on those same
grounds. We took this to indicate that the protection from
$\ord{\lqcd/m_{c,b}}$-corrections at zero recoil that Luke's theorem
provides appears to extend over the full range of recoils. It also
meant that two independent determinations of the Isgur-Wise function
could be obtained from our results for $h^+$ and $h_{A_1}$. The two
Isgur-Wise functions found in this way were
identical indicating that the spin component of heavy quark
symmetry is nearly unbroken in this particular situation.  This procedure
for obtaining the Isgur-Wise function was repeated for two values of
the light-quark mass: $m_q=0$ and $m_q=m_s$. The slopes or the
corresponding Isgur-Wise functions, $\xi_{u,d}$ and $\xi_s$, are
at $\w=1$:
\beq
\label{xiudp}
\xi_{u,d}'|_{w=1}=-\l[0.9\er{2}{3}\stat\er{4}{2}\syst\r]
\eeq
and
\beq
\label{xisp}
\xi_s'|_{w=1}=-\l[1.2\er{2}{2}\stat\er{2}{1}\syst\r]
\ .\eeq
We were also able to extract some of the form factors which appear
at order $1/m_{b,c}$ in the description of the semi-leptonic decays,
thereby probing some of the more intricate details of HQET. 

When studying the radiative decay $B\to K^*\g$ we found that ${\hat
T}_2$ defined in \eq{t2hat} suffered power corrections on the order of
10\% at $m_b$ and 30\% at $m_c$, very much in line with leptonic decays. 
This information, however,
does not tell us how the process at $q^2{=}0$ 
scales with the mass of the initial
heavy quark. To determine this scaling one has to understand the
$q^2$ dependence of at least one of the relevant form factors as discussed
in \sec{bsgsec}.
Since the precision of the lattice prediction is vitally
dependent on understanding this scaling, it is important to settle
this issue soon.

It is also important, in the near future, to improve calculations of
the $B$-parameter for $B-\bar B$ mixing and to perform systematic
studies of semi-leptonic $B\to\rho(\pi)\ell\bar\nu$. These processes
can be approached with the methods described above and are important for
determining the CKM matrix elements $|V_{td}|$, $|V_{ts}|$ and
$|V_{ub}|$, respectively. To permit an alternative determination of
$|V_{cb}|$, the UKQCD Collaboration is furthermore undertaking a study
of the semi-leptonic decays of baryons, such as the $\Lambda_b$, which
contain one heavy quark and two light quarks.  In the process, 
many more tests of heavy-quark symmetry will be performed. Some time
and effort should also be spent on non-leptonic decays.  $B\to J/\Psi
K_s$ or $B\to \pi^+\pi^-$ decays, for instance, are important for
understanding CP violation in the bottom quark sector.  The stumbling
blocks here are both technical and conceptual.  Technically, present
day lattices are just too small to separate the hadrons in the final
state and conceptually, how to determine the relevant final-state
phase shifts is not yet understood\cite{bersonmaites}.

\medskip
The results presented here are, for the most part, products of a first
generation of calculations. The errors quoted on these quantities will
therefore decrease in the months and years to come as we control
better the various sources of systematic uncertainties and design more
efficient lattice actions and faster algorithms. When working with
heavy quarks, aside from un-quenching of course, the challenge will be
to reduce discretization errors. This will involve extrapolating
results to the continuum limit by performing high statistics
simulations on lattices with different of lattice spacings. It will
also involve designing actions, such as the Clover or even perfect
actions mentioned above, which converge more rapidly to the continuum
limit. One further expects effective theories, such as NRQCD or the
lattice variant of HQET, to play a growing role in the description of
heavy-quark decays.  In any event, Lattice QCD has already shown
itself to be a valuable quantitative tool for strong interaction
physics and it will undoubtedly provide increasingly precise and 
varied results in the future.

\section{Acknowledgments}
I would like to thank the organizers of the XXXIV$^{th}$ Cracow School
of Theoretical Physics in Zakopane, Poland, for their hospitality and
for having put together a very enjoyable and stimulating session.  I
would also like to thank Jonathan Flynn, Henning Hoeber, Juan Nieves,
Chris Sachrajda, Nicoletta Stella, Hartmut Wittig and the rest of the UKQCD
Collaboration for very stimulating discussions on different aspects of
the results presented here. I am also grateful to Tanmoy Bhattacharya,
Philippe Boucaud and Brian Gough for
sharing with me the latest results for $\bksg$ of their respective groups.

\vfill



\end{document}